

\magnification\magstep1
\baselineskip14pt
\vsize23.0truecm 

\input miniltx
\input graphicx


\font\csc=cmcsc10
\font\bigbf=cmbx12 
\font\smallrm=cmr8 
\def\hop{\smallskip\noindent}

\def\hatt{\widehat}
\def\Pr{{\rm Pr}}
\def\half{\hbox{$1\over2$}}

\def\sumin{\sum_{i=1}^n}
\def\prodin{\prod_{i=1}^n}
\def\dell{\partial} 
\def\rootn{\sqrt{n}}
\def\square{{\ \vrule height0.5em width0.5em depth-0.0em}}
\def\sd{{\rm sd}}
\def\var{{\rm var}}  
\def\truth{{\rm truth}}
\def\model{{\rm model}}
\def\true{{\rm true}}
\def\ave{{\rm ave}}
\def\theory{{\rm theory}} 
\def\diag{{\rm diag}}
\def\niente{\phantom{A}}

\def\baselineskiptables{\baselineskip8pt}

\def\tilda{\widetilde}
\def\N{{\rm N}}
\def\E{{\rm E}}
\def\RR{\mathord{I\kern-.3em R}}
\def\Var{{\rm Var}}
\def\arr{\rightarrow} 
\def\midd{\,|\,}
\def\tr{{\rm t}}
\def\Tr{{\rm Tr}}
\def\mtrix{\pmatrix}
\def\eps{\varepsilon} 

\def\ML{{\rm ML}}
\def\PL{{\rm PL}}
\def\QL{{\rm QL}}
\def\lik{{\rm l}}
\def\pl{{\rm pl}}
\def\ql{{\rm ql}}
\def\section{\bigskip}
\def\subsection{\medskip}

\def\fermat#1{\setbox0=\vtop{\hsize4.00pc
        \smallrm\raggedright\noindent\baselineskip9pt
        \rightskip=0.5pc plus 1.5pc #1}\leavevmode
        \vadjust{\dimen0=\dp0
        \kern-\ht0\hbox{\kern-4.00pc\box0}\kern-\dimen0}}
        
\bigskip
\centerline{\bigbf ML, PL, QL in Markov chain models} 

\smallskip
\centerline{\bf Nils Lid Hjort and Cristiano Varin}

\smallskip
\centerline{\bf University of Oslo and University of Udine}

\smallskip
\centerline{\sl -- April 2005 --} 

\medskip
{\smallskip\narrower\noindent\baselineskip12pt 
{\csc Abstract.} 
In many spatial and spatial-temporal models, 
and more generally in models with complex dependencies,
it may be too difficult to carry out full maximum
likelihood (ML) analysis. Remedies include 
the use of pseudo-likelihood (PL) and quasi-likelihood (QL)
(also called the composite likelihood). 
The present article studies the ML, the PL and the QL
methods for general Markov chain models,
partly motivated by the desire to understand the 
precise behaviour of PL and QL methods in settings
where this can be analysed. We present limiting 
normality results and compare performances 
in different settings. The PL and QL methods can be
seen as maximum penalised likelihood methods. 
We find that the QL strategy is typically 
preferable to the PL, and that it loses very little
to the ML, while earning in model robustness. 
It has also appeal and potential as a modelling tool. 
Our methods are illustrated for analysis 
of DNA sequence evolution type models. 

\smallskip\noindent
{\csc Key words:} \sl 
composite likelihood, 
DNA, 
estimator performance, 
Markov chains, 
pseudo-likelihood, 
quasi-likelihood,
spatial models 
\smallskip}

\section
\centerline{\bf 1. Introduction and summary}

\hop
In various spatial and spatial-temporal statistical models 
the direct use of the full maximum likelihood (ML) method 
is often cumbersome or even too difficult to carry 
out numerically. There are also situations where the
ML method can be expected to rely too much on the
correctness of the model used, i.e.~the method 
may not be robust. These practical and theoretical 
aspects have motivated constructions of various alternative 
estimation methods for different situations. 
Some of these still keep the likelihood flavour, 
specifically the pseudo-likelihood (PL) method, 
going back to papers by Besag (1974, 1977), 
and the quasi-likelihood or composite likelihood (QL) method, 
stemming from Hjort and Mohn (1987), Hjort and Omre (1994),
and Lindsay (1988). 

The PL-likelihood is a product over conditional 
densities, where the conditioning in question may
be with respect to a local neighbourhood or a
larger set. In Markov models, the conditional distribution
at a site given neighbourhood information 
is identical to the distribution at the site 
given data from all other sites, making
the PL particularly attractive. The QL-likelihood,
on the other hand, is typically a product 
over all bivariate distributions, for all sets
of neighbours, or for pairs of sites that are 
farther apart. The QL comes in several forms, 
like in pairwise and triplewise fashion, and is
particularly suited to situations where the bivariate
or trivariate marginal distributions are easier
to handle than the high-dimensional distribution
of the full data set. Hjort and Omre (1994) 
suggested using QL for inference in traditional 
geostatistical models, for example for estimating
parameters of variograms. Glasbey (2001) models solar 
radiation data via a class of non-linear autoregressive 
time series, and carries out estimation using 
the QL method; he focusses on joint marginal distributions
at low lags, using multivariate normal mixtures for these. 
Heagerty and Lele (1998) exploited the same idea to binary spatial data 
and Varin, H\o st and Skare (2005) to arbitrary
generalised linear geostatistical models. 
Renard, Molenberghs and Geys (2004) considered 
the QL for multilevel probit models, while Henderson 
and Shimakura (2003) used it for frailty models 
for longitudinal count data. Other possible 
applications are image analysis (Nott and Ryd\'en, 1999), 
genetics (Fearnhead and Donnelly, 2002),  
multivariate survival analysis (Parner, 2001), 
and modelling and inference for mixed ordinal and continuous data 
(de Leon, 2004). A general recent discussion on theoretical aspects 
and possible applications of QL is given in Cox and Reid (2004). 

\subsection
{\sl 1.1. Estimation in Markov chains.} 
Our present aim is to discuss and compare these three methods  
of estimation, the ML, the PL and the QL, 
in the case of traditional Markov chain models. 
In such models the ML would typically 
be the method of choice, on the merits of simplicity 
and efficiency under the assumed model, but the PL and QL 
alternatives are still of interest, with their own 
relative advantages. For example, it is explained
below that the QL method performs rather similarly 
to the full ML method, but that it invests some extra 
efforts in estimation the model so as to better 
achieve closeness of the inferred estimated 
equilibrium distribution to the real one. Similarly,
the PL makes it a priority to estimate well 
the conditional distributions of observations given
neighbouring information. These `extraneous aspects'
of the model fitting work might not be sufficiently well
taken care of by the ML method, in cases where the
model used is not perfect. Thus we find, in Sections 3 and 4,
that the QL and PL methods act as maximum penalised 
likelihood strategies.

Analysing Markov chain models with model-robust strategies
has independent interest in view of the broad and expanding 
horizon of applications of such models. 
We have also been motivated
by the desire to understand better the behaviour 
and performance of PL and QL methods, in a setting
where they may actually be analysed precisely;
in more complex spatial or spatial-temporal models
it is much harder to reach clear results. 
Thus there is also a pedagogical element in our exposition,
in that finite-state stationary Markov chains 
are among the best-studied models, down to the 
statistics undergraduate level. 

To establish the required notation, 
let $X_0,X_1,\ldots$ be an irreducible Markov chain 
on a finite state space with stationary transition probabilities. 
A parametric model is considered, of the form
$$\pi_{a,b}=\Pr\{X_i=b\midd X_{i-1}=a\}=p_{a,b}(\theta)
  \quad {\rm for\ }a,b=1,\ldots,S, \eqno(1.1)$$
where $\theta$ is some underlying parameter vector
and $S$ the number of states. 
Our agenda is to assess and compare various estimation methods
for $\theta$ (and hence for transition probabilities
$p_{a,b}(\theta)$, and for other functions thereof). 
The chain is observed from time zero to time $n$, 
and we will examine estimation methods from a large-sample
perspective in which $n$ grows towards infinity. 
We note that the special case of no parametric assumptions 
at all, corresponding to all $p_{a,b}$s being unknown,  
is subsumed in the general apparatus as the special case 
of $S(S-1)$ parameters $p_{a,b}$, with $b=1,\ldots,S-1$ and 
$a=1,\ldots,S$. 

The traditional method is the ML or maximum likelihood method,
consisting in maximising 
$$\lik_n(\theta)=\prod_{i=1}^n \Pr_\theta\{X_i=x_i\midd X_{i-1}=x_{i-1}\}
        =\prod_{a,b} p_{a,b}(\theta)^{N_{a,b}}. \eqno(1.2)$$
Here the Markov assumption is of course used, and 
$N_{a,b}=\sum_{i=1}^n I\{(X_{i-1},X_i)=(a,b)\}$
counts the number of transitions from $a$ to $b$. 
Its properties are well understood and reported on in
the literature, and are briefly reviewed in Section~2. 
The PL or maximum pseudo-likelihood method, on the other hand, 
maximises 
$$\pl_n(\theta)=\prod_{i=1}^{n-1}\Pr_\theta\{X_i=x_i
        \midd{\rm rest\ of\ data}\}. $$
Here 
$$\Pr\{X_i=b\midd X_{i-1}=a,X_{i+1}=c\}
        ={\Pr\{X_{i-1}=a\}\,p_{a,b}(\theta)p_{b,c}(\theta)
         \over \Pr\{X_{i-1}=a\}\,p_{a,c}^{(2)}(\theta)}, $$
with the stationary probability $\Pr\{X_{i-1}=a\}$ cancelling,
so that in fact 
$$\pl_n(\theta)=\prod_{i=1}^{n-1}\Pr_\theta\{X_i=x_i
        \midd X_{i-1}=x_{i-1},X_{i+1}=x_{i+1}\} $$
with dependence only upon the nearest neighbours. 
Here $p_{a,c}^{(2)}(\theta)=\sum_b p_{a,b}(\theta)p_{b,c}(\theta)$ 
are the two-step transition probabilities. 
This leads to 
$$\pl_n(\theta)=\prod_{a,b,c}\Bigl\{{p_{a,b}(\theta)p_{b,c}(\theta)
        \over p_{a,c}^{(2)}(\theta)}\Bigr\}^{N_{a,b,c}}, \eqno(1.3)$$
where $N_{a,b,c}=\sum_{i=1}^{n-1} I\{(X_{i-1},X_i,X_{i+1})=(a,b,c)\}$
counts the number of $(a,b,c)$ transitions. 
Besag's (1974, 1977) original PL ideas were 
similar to the above, but in a context of two-dimensional
Markov fields. 

The method just described may be thought of as a first-order 
\PL{} method. Also second-order and higher-order pseudo-likelihoods 
can be considered. For example, 
$$\Pr\{X_i=b,X_{i+1}=c\midd X_{i-1}=a,X_{i+2}=d\}
        ={p_{a,b}(\theta)p_{b,c}(\theta)p_{c,d}(\theta)
        \over p_{a,d}^{(3)}(\theta)}, $$
which suggests using 
$$\eqalign{\pl_{n,2}(\theta)
&=\prod_{i=1}^{n-2}\Pr_\theta
        \{X_i=x_i,X_{i+1}=x_{i+1}\midd{\rm rest\ of\ data}\} \cr
&=\prod_{a,b,c,d}\Bigl\{
        {p_{a,b}(\theta)p_{b,c}(\theta)p_{c,d}(\theta)
        \over p_{a,d}^{(3)}(\theta)}\Bigr\}^{N_{a,b,c,d}}. \cr}$$

The third class of estimators we consider is that of maximising 
the \QL{} function, defined as 
$$\ql_n(\theta)=\prod_{i=1}^n\Pr_\theta\{X_{i-1}=x_{i-1},X_i=x_i\}
        =\prod_{a,b}\{p_a(\theta)p_{a,b}(\theta)\}^{N_{a,b}}. \eqno(1.4)$$
Here we assume, for concreteness and simplicity, that the chain 
starts out in its equilibrium distribution, so that 
$\Pr_\theta\{X_i=a\}=p_a(\theta)$, determined by the 
transition probabilities $p_{a,b}(\theta)$. 
This assumption is not vital to our conclusions, however,
mainly because $\Pr\{X_i=a\}$ regardless of starting conditions
converges with exponential quickness towards $p_a(\theta)$. 

The \QL{} (or composite likelihood) idea is more generally 
to form products of marginal likelihoods for many small subsets. 
The third-order version, for example, would be 
$$\eqalign{\ql_{n,3}(\theta)
&=\prod_{i=1}^{n-1}\Pr_\theta\{X_{i=1}=x_{i-1},
        X_i=x_i,X_{i+1}=x_{i+1}\} \cr
&=\prod_{a,b,c}\{p_a(\theta)p_{a,b}(\theta)
   p_{b,c}(\theta)\}^{N_{a,b,c}}.\cr}$$
Below we explore general versions of both PL and QL. 
Versions of the \QL{} were first discussed in
Hjort and Mohn (1987) in connection with analysis
of satellite data, and then proposed for geostatistical
models in Hjort and Omre (1994); the QL is particularly
suited to models where where lower-dimensional marginal
distributions (e.g.~bivariate and trivariate) 
are rather simpler to handle than the full likelihood. 
See also Hjort and Mostad (1998), where the ideas 
were used for spatial point processes;
there the simultaneous distributions required 
for ML analysis are often very difficult to handle. 

\subsection
{\sl 1.2. The present paper.} 
The layout of our article is as follows. 
We reach precise limiting normality results
for the ML, the QL and the PL methods in
Sections 2, 3, 4, exhibiting in each case formulae
for the limiting variance matrices. The ML results
are classic and date back to at least 
Anderson and Goodman (1957) and Billingsley (1961a, 1961b), 
but the QL and PL results appear to be new. 
We choose to deal with the QL before the PL since 
the latter is more cumbersome. 
Our results are then used in Section~5 to go through 
a generous list of special Markov chain models, where we can 
analyse and compare performances. Simulation experiments
are carried out to compare the asymptotic descriptions
with actual finite-sample precision. In Section 6
the methods are illustrated in connection with analysis
of DNA sequence evolution type models, and in Section 7 
we discuss and pinpoint differences between the methods,
in cases where the assumed parametric model is 
not entirely correct. Our paper ends with a list 
of concluding remarks. Among the points we make is
that the good aspects of the QL method should be 
actively used on the statistical modelling front. 

\section
\centerline{\bf 2. The ML method}

\hop 
We work in this section under the conditions of the 
postulated Markov chain model, so that (1.1) is in
force for a certain $\theta_0$. 
Let $u_{a,b}(\theta)$ (a $p$-vector) and $i_{a,b}(\theta)$ 
(a $p\times p$-matrix) be the first and second derivatives of 
$\log p_{a,b}(\theta)$. Introduce
$$J_a=\sum_b p_{a,b}u_{a,b}(\theta_0)u_{a,b}(\theta_0)^\tr
  \quad {\rm and} \quad 
  J=\sum_a p_aJ_a. \eqno(2.1)$$
We shall on occasion omit the argument $\theta_0$
from the notation, as here, for brevity of presentation. 
Taking two successive $\theta$-derivatives of the equation
$\sum_b p_{a,b}(\theta)=1$ gives the identities
$\sum_b p_{a,b}(\theta)u_{a,b}(\theta)=0$ and 
$J_a=-\sum_b p_{a,b}(\theta_0)i_{a,b}(\theta_0)$,
so that also 
$J=-\sum_{a,b}p_a(\theta_0)p_{a,b}(\theta_0)i_{a,b}(\theta_0)$. 
For some information calculations 
it is also practical to note that 
$$J_a=\sum_b {1\over p_{a,b}}{\dell p_{a,b}\over \dell\theta}
   \Bigl({\dell p_{a,b}\over \dell\theta}\Bigr)^\tr. \eqno(2.2)$$

Now focus on the \ML{} estimator $\hatt\theta$, 
which under regularity conditions is the solution to 
$$U_n(\theta)=\dell\log\lik_n(\theta)/\dell\theta
        =\sum_{a,b} N_{a,b}u_{a,b}(\theta)=0. $$
We may now state and prove the following. 
It is assumed that the usual type regularity conditions
are in force, including smoothness of the second order
derivatives of the transition probability functions;
also, the true parameter point $\theta_0$ is taken
to be an inner point of the parameter space. 
More details and discussion regarding regularity conditions 
can be found in e.g.~Billingsley (1961a, 1961b). 

{\smallskip\sl
{\csc ML Proposition.}
Under model conditions,
$\rootn(\hatt\theta-\theta_0)\arr_d\N(0,J^{-1})$. 
\smallskip}

{\csc Sketch of proof.}
Standard arguments, using one-step Taylor analysis
from $U_n(\hatt\theta)=0$, show that 
$\rootn(\hatt\theta-\theta_0)
        \approx \tilda J_n^{-1}U_n(\theta_0)/\rootn$, where 
$$\tilda J_n=-n^{-1}{\dell^2\log{\rm l}(\theta_0)
   \over \dell\theta\dell\theta^\tr}
        =-n^{-1}\sum_{a,b}N_{a,b}i_{a,b}(\theta_0). $$
Here $\tilda J_n$ is seen to tend in probability to $J$. 
Furthermore, 
$$U_n(\theta_0)/\rootn=\sum_{a,b}\rootn
   (N_{a,b}/n-p_a(\theta_0)p_{a,b}(\theta_0))u_{a,b}(\theta_0)
        \arr_d \sum_{a,b}Z_{a,b}u_{a,b}(\theta_0), $$
where the $Z_{a,b}$s form a random zero-mean normal vector,
the limiting distribution of the the individual 
$\rootn(N_{a,b}/n-p_a(\theta_0)p_{a,b}(\theta_0)$. 
Their covariances can be worked out to be 
$${\rm cov}(Z_{a,b},Z_{c,d})=p_ap_{a,b}(\delta_{a,c}\delta_{b,d}-p_cp_{c,d})
        +p_{a,b}p_{c,d}(p_a\gamma_{b,c}+p_c\gamma_{d,a}), \eqno(2.3)$$
in which 
$$\gamma_{a,b}=\sum_{k=0}^\infty(p_{a,b}^{(k)}-p_b). \eqno(2.4)$$
See for example Basawa and Rao (1980, Ch.~4). 
The $\gamma_{a,b}$ quantities are finite due to exponential convergence 
of $p_{a,b}^{(k)}$ to $p_b$.  

We need to work out an expression for 
$$\Var\Bigl(\sum_{a,b}Z_{a,b}u_{a,b}\Bigr)
        =\sum_{a,b}\sum_{c,d}{\rm cov}(Z_{a,b},Z_{c,d})u_{a,b}u_{c,d}^\tr. $$
Upon using the fact that $\sum_b p_{a,b}u_{a,b}=0$, 
even several times, we find that this matrix is simply $J$. 
In conclusion, therefore, 
$\rootn(\hatt\theta-\theta)$ 
must tend in distribution to $J^{-1}\N(0,J)=\N(0,J^{-1})$,
as was to be proved. 
\square

\smallskip
{\csc Remark.} 
Note that the $J$ matrix that governs the ML precision
ended up not depending on the $\gamma_{a,b}$ quantities,
even though they technically speaking were used in the proof.
This is connected to the fact that 
$$\rootn(N_{a,b}/N_{a,\cdot}-p_{a,b})
  \arr_d p_a^{-1}(Z_{a,b}-p_{a,b}Z_{a,\cdot}), $$
and these have covariance depending only on 
the $p_a$s and $p_{a,b}$s. The $\gamma_{a,b}$ quantities
will however be involved in the limit distributions
for both the QL and the PL methods. 

\section
\centerline{\bf 3. The QL method}

\hop 
For the \QL{} of (1.4), 
$$\log\ql_n(\theta)=\sum_a N_{a,\cdot}\log p_a(\theta)
        +\sum_{a,b}N_{a,b}\log p_{a,b}(\theta), $$
where $N_{a,\cdot}=\sum_b N_{a,b}$; similar notation will be 
used below. Thus the log-ql function is the sum of the ordinary
log-likelihood and an additional term that aims at making 
the $p_a(\theta)$s come close to the observed $N_{a,\cdot}/n$. 

The generalisation to higher-order quasi-likelihoods is 
also instructive. For the third-order version, for example, 
$$\log\ql_{n,3}(\theta)=\sum_a N_{a,\cdot,\cdot}\log p_a(\theta)
        +\sum_{a,b}N_{a,b,\cdot}\log p_{a,b}(\theta)
        +\sum_{b,c}N_{\cdot,b,c}\log p_{b,c}(\theta). $$
Here it is seen that 
$N_{a,b,\cdot}$ and $N_{\cdot,a,b}$ are practically identical;
they differ at most 1 from $N_{a,b}$. 
In the large-sample framework there is therefore 
no difference between $\rootn(N_{a,b}/n-p_ap_{a,b})$
and sister versions that employ $N_{a,b,\cdot}$ or $N_{\cdot,a,b}$. 
Thus, in general, with a fixed order $k$ for our \QL{} method, 
$$\log{\rm ql}_k(\theta)=\sum_a N_a\log p_a(\theta)
        +(k-1)\sum_{a,b}N_{a,b}\log p_{a,b}(\theta), \eqno(3.1)$$
to a good approximation, as $n$ grows. 
We note that in cases where the equilibrium distribution 
is independent of $\theta$, as in some symmetric models, 
for example, the \QL{} method is practically equivalent 
to the classic \ML{} method. 

To explore its asymptotic performance, 
introduce $v_a(\theta)=\dell\log p_a(\theta)/\dell\theta$,
along with 
$$H=\sum_a p_av_av_a^\tr, \quad
  G=\sum_{a,b}p_a\bar\gamma_{a,b}v_av_b^\tr, \quad
  L=\sum_{a,b}p_ap_{a,b}u_{a,b}\kappa_b^\tr, \eqno(3.2)$$
which involve further quantities
$$\kappa_b=\sum_c\gamma_{b,c}v_c=\sum_{k\ge0}\sum_c(p_{b,c}^{(k)}-p_c)v_c
  \quad {\rm and} \quad 
  \bar\gamma_{a,b}=\sum_{k\ge1}(p_{a,b}^{(k)}-p_b). \eqno(3.3)$$
The various quantities here are computed under 
the true parameter $\theta_0$. When comparing definitions 
of $\gamma_{a,b}$ and $\bar\gamma_{a,b}$
we see that $\bar\gamma_{a,b}=\gamma_{a,b}-(\delta_{a,b}-p_b)$. 
Also note that $H=-\sum p_a\dell^2\log p_a/\dell\theta\dell\theta^\tr$. 
We may now formulate the main result on QL estimation. 
In order not to overburden the notation we allow 
the $k$th order QL estimator to be denoted $\hatt\theta$ 
in this section, the one maximising $\log\ql_{n,k}(\theta)$ of (3.1);
it should not lead to confusion with the ML estimator 
of the previous section. 

{\smallskip\sl
{\csc QL Proposition.}
Under model conditions, 
$\rootn(\hatt\theta-\theta_0)\arr_d\N(0,J_k^{-1}K_kJ_k^{-1})$, where 
$$J_k=(k-1)J+H
  \quad {\sl and} \quad 
  K_k=(k-1)^2J+H+G+G^\tr+(k-1)(L+L^\tr). $$
\smallskip}

{\csc Sketch of proof.}
We employ arguments similar to those utilised in Section 2. 
The key to understanding $\rootn(\hatt\theta-\theta)$
there was the Taylor-based approximation 
$\tilda J_n^{-1}U_n(\theta)/\rootn$. 
Here the same technique and its consequence take the form 
$$\rootn(\hatt\theta-\theta_0)\approx J_{n,k}^{-1}
        {1\over \rootn}{\dell\ql_{n,k}(\theta_0)\over \dell\theta}
        \arr_d \bar J_k^{-1}\N(0,\bar K_k)
    =\N(0,\bar J_k^{-1}\bar K_k\bar J_k^{-1}) \eqno(3.4)$$
with the appropriate matrices $\bar J_k$ and $\bar K_k$,
and the rest of the proof consists in showing that 
indeed $\bar J_k$ and $\bar K_k$ are the $J_k$ and $K_k$ 
matrices given in the proposition. 

Here $\bar J_k$ is the limit in probability of 
$J_{n,k}$, minus $n^{-1}$ times the second derivative matrix 
of $\log{\rm ql}_k(\theta_0)$. That is, 
$$\bar J_k=-\sum_a p_a{\dell^2\log p_a\over \dell\theta\dell\theta^\tr}
        -(k-1)\sum_{a,b}p_ap_{a,b}i_{a,b}
        =H+(k-1)J, $$
which agrees with the claim made. 
The other ingredient of the limit distribution is $\bar K_k$, 
the limiting covariance matrix of 
$$\sum_a\rootn\Bigl({N_a\over n}-p_a\Bigr)v_a
   +(k-1)\sum_{a,b}\rootn\Bigl({N_{a,b}\over n}
   -p_ap_{a,b}\Bigr)u_{a,b}. $$
By results used in Section 2 it may be expressed as 
$$\bar K_k=\Var\Bigl\{\sum_a Z_{a,\cdot}v_a
   +(k-1)\sum_{a,b}Z_{a,b}u_{a,b}\Bigr\}. $$
Some algebraic work, using (2.3) repeatedly, 
verifies that 
$$
\Var\Bigl(\sum_a Z_{a,\cdot}v_a\Bigr)=H+G+G^\tr
\quad {\rm while} \quad
\E\sum_{a,b,c} Z_{a,b}Z_{c,\cdot}u_{a,b}v_c^\tr=L. $$
This shows that $\bar K_k$ indeed equals the $K_k$ matrix
given above.
\square 

\smallskip
{\csc Remark 3.1.} 
It is clear that `the QL method' is really a class
of methods, the simplest ones being the pairwise,
the triplewise and the quadruplewise. If the order 
$k$ is allowed to grow, we learn from the QL 
proposition that $J_k^{-1}K_kJ_k^{-1}$
comes closer to $J^{-1}$, i.e.~the $k$th-order QL
will perform almost equivalently to the ML method.
\square 

\def\pen{{\rm pen}}

\smallskip
{\csc Remark 3.2.} 
We see from (3.1) that the QL estimators maximise
$$\log\lik_n(\theta)+(k-1)^{-1}\pen_{Q,n}(\theta), \eqno(3.5)$$ 
where $\pen_{Q,n}(\theta)$ is a term that is large when 
all marginal probabilities $p_a(\theta)$ are 
well estimated. Thus the QL may be seen as 
a maximum penalised likelihood method, with 
$(k-1)^{-1}$ as penalisation constant; again, 
if $k$ grows, then the QL becomes equivalent to the ML. 
The (3.5) view of the QL even invites Bayesian or
empirical Bayesian connotations; the method
is similar to a Bayesian method with a prior 
or an empirically estimated prior for the marginal distribution. 
In some detail, suppose one uses the prior distribution
$$g(\theta)\propto\exp\Bigl\{-\rho\sum_a p_a^0
   \log{p_a^0\over p_a(\theta)}\Bigr\} $$
for some $p_a^0$ values, corresponding to the 
belief that the equilibrium distribution will 
tend to be close to $(p_1^0,\ldots,p_S^0)$,
and with larger values of $\rho$ when the prior belief
is strong. An empirical version of this inserts 
$N_{a,\cdot}/n$ for $p_a^0$. The QL method can then 
be seen as a maximum posterior density method, 
with the strong $\rho$ value $n/(k-1)$. 
\square 

\smallskip
{\csc Remark 3.3.}
In models where the equilibrium distribution
does not vary much with $\theta$, the $v_a(\theta)$s
will contribute little, making the $G$, $H$, $L$
matrices small in size, and thence the QL close 
to the ML in performance. The particular case 
of the equilibrium unperturbed by $\theta$ 
corresponds to exact equivalence. 
\square 

\section 
\centerline{\bf 4. The PL method}

\hop 
This method is more cumbersome to analyse than the other two,
in that two-step or higher-order step probabilities unavoidably
enter the calculations. We have 
$$\log\pl_n(\theta)
=\sum_{a,b}N_{a,b,\cdot}\log p_{a,b}(\theta)
        +\sum_{b,c}N_{\cdot,b,c}\log p_{b,c}(\theta)
        -\sum_{a,c}N_{a,\cdot,c}\log p_{a,c}^{(2)}(\theta). \eqno(4.1)$$  
Note that $N_{a,b,\cdot}$ and $N_{\cdot,a,b}$ are very close,
differing with at most 1, so the two first terms 
here add up to $2\sum_{a,b}N_{a,b}\log p_{a,b}(\theta)$ 
as far as all large-sample analysis is concerned. 

The limit distribution is again found via arguments similar 
to those used in previous sections. In order to accurately 
describe this limit distribution we need to introduce
several matrices. In addition to 
$u_{a,b}=\dell\log p_{a,b}/\dell\theta$, define 
$$w_{a,c}=\dell\log p_{a,c}^{(2)}/\dell\theta
   =\sum_b {p_{a,b}p_{b,c}\over p_{a,c}^{(2)}}(u_{a,b}+u_{b,c}), $$
along with 
$$\eqalign{
M&=\sum_a p_aM_a=\sum_{a,c}p_ap_{a,c}^{(2)}w_{a,c}w_{a,c}^\tr, \cr
Q&=\sum_{a,c,d,f}p_ap_{a,d}p_{d,c}p_{c,f}w_{a,c}w_{d,f}^\tr, \cr
R&=\sum_{a,b,c}p_ap_{a,b}p_{b,c}(u_{a,b}+u_{b,c})w_{a,c}^\tr. \cr}
  \eqno(4.2)$$
In fact $M=R$, which can be seen starting from 
$$p_{a,c}^{(2)}w_{a,c}={\dell\over \dell\theta}p_{a,c}^{(2)}
   =\sum_b{\dell\over \dell\theta}p_{a,b}p_{b,c}
   =\sum_bp_{a,b}p_{b,c}(u_{a,b}+u_{b,c}). $$
We finally note, before coming to the main result
about PL estimation, that 
$$M_a=-\sum_cp_{a,c}^{(2)}{\dell^2\log p_{a,c}^{(2)}
   \over \dell\theta\dell\theta^\tr}
   =\sum_c{1\over p_{a,c}^{(2)}}
    {\dell p_{a,c}^{(2)}\over \dell\theta}
    \Bigl({\dell p_{a,c}^{(2)}\over \dell\theta}\Bigr)^\tr. $$
In this section $\hatt\theta$ is the PL estimator,
maximising $\log\pl_n(\theta)$ of (4.1). 

{\smallskip\sl
{\csc PL Proposition.}
Under model conditions, 
$\rootn(\hatt\theta-\theta_0)\arr_d \N(0,J_0^{-1}K_0J_0^{-1})$, 
where $J_0=2J-M$ and 
$$K_0=4J+M+Q+Q^\tr-2(R+R^\tr)=4J-3M+Q+Q^\tr. $$
\smallskip}

{\csc Sketch of proof.} 
The programme is to verify that 
$$\rootn(\hatt\theta-\theta_0)
   \doteq_d J_{0,n}^{-1}n^{-1/2}\dell\log\pl_n(\theta_0)/\dell\theta
   \arr_d J_0^{-1}\N(0,K_0), $$
where $J_0$ is the limit in probability of 
$J_{0,n}=-n^{-1}\dell^2\log\pl_n(\theta_0)/\dell\theta\dell\theta^\tr$
and with $K_0$ the appropriate variance matrix 
of the limiting distribution of 
$n^{-1/2}\dell\log\pl_n(\theta_0)/\dell\theta$. We find 
$$J_{0,n}=-2\sum_{a,b}{N_{a,b}\over n}
  {\dell^2\log p_{a,b}(\theta)\over \dell\theta\dell\theta^\tr}
  +\sum_{a,c}{N_{a,\cdot,c}\over n}
  {\dell^2\log p_{a,c}^{(2)}\over \dell\theta\dell\theta^\tr}
  \arr_p 2J-M $$
as required, since $N_{a,b}/n\arr_p p_ap_{a,b}$
and $N_{a,\cdot,c}/n\arr_p p_ap_{a,c}^{(2)}$, and because of the 
likelihood-related identity for $J_a$ mentioned
after (2.1), along with the similar 
$$M_a=\sum_bp_{a,b}^{(2)}w_{a,b}w_{a,b}^\tr
   =-\sum_bp_{a,b}{\dell\log p_{a,b}^{(2)}\over \dell\theta\dell\theta^\tr}. $$
Note also that $\sum_bp_{a,b}^{(2)}w_{a,b}=0$, just as
$\sum_bp_{a,b}u_{a,b}=0$. 

For the second key matrix, start with the collection 
of $Z_{a,b,c}$ zero-mean normal variables which are 
the simultaneous limits in distribution of the variables
$\rootn(N_{a,b,c}/n-p_ap_{a,b}p_{b,c})$, and for which 
the covariance structure is worked out in this article's
Appendix. We have 
$$\eqalign{
n^{-1/2}{\dell\log\pl_n(\theta_0)\over \dell\theta}
&=2\sum_{a,b}\rootn\Bigl({N_{a,b}\over n}-p_ap_{a,b}\Bigr)u_{a,b}
   -\sum_{a,c}\rootn\Bigl({N_{a,\cdot,c}\over n}
   -p_ap_{a,c}^{(2)}\Bigr)w_{a,c} \cr
&\arr_d 2\sum_{a,b}Z_{a,b}u_{a,b}-\sum_{a,c}Z_{a,\cdot,c}w_{a,c}
   =2A-B, \cr}$$
say, where $Z_{a,\cdot,c}=\sum_b Z_{a,b,c}$,  
along with $Z_{a,b}=Z_{a,b,\cdot}=\sum_c Z_{a,b,c}$. 
Here $A$ has variance matrix $J$, as seen in Section 2. 
The proof of the proposition is completed by showing 
that the variance matrix of $B$ is $M+Q+Q^\tr$,
utilising result (A.2) of the Appendix, and that 
$\E(AB^\tr)=R$, which involves both covariance results
(A.1) and (A.2). \square 

\smallskip
{\csc Remark.} 
We see from (4.1) and the ensuing discussion that the PL
method essentially aims at maximising 
$$\log\lik_n(\theta)-\half\pen_{P,n}(\theta), \eqno(4.3)$$ 
where the penalty term is 
$\sum_{a,c}N_{a,\cdot,c}\log p_{a,c}^{(2)}(\theta)$.
Thus, as with the  the QL method, the PL can be 
seen as a maximum penalised likelihood strategy. 
One may argue that the penalisation term (3.5) 
for the QL method is rather more natural and fruitful 
than (4.3) for the PL, however. A large value of 
$\pen_{P,n}(\theta)$ is advantageous for fitting 
two-step probabilities, but the PL risks encouraging
a small value of this term. This suggests that the QL 
might have performance and robustness advantages 
over the PL. This will be confirmed in our case studies;
see in this connection also Section 7. 
\square 

\section
\centerline{\bf 5. Examples and illustrations}

\hop
In this section we go through a fair list of 
Markov chain models, where we may examine 
and compare the behaviour of ML, PL, QL methods. 
Studying the QL method involves finding matrices
$G$, $H$, $L$, while precision of the PL method 
is decided by further matrices $M$, $Q$.
It may be difficult to find explicit formulae
for these, when the model studied is complicated
with more than a few parameters, but one may still
carry out precise numerical evaluations and 
comparisons of limit distributions, at each
point or region of interest in the parameter space,
by computing the matrices numerically. 


\subsection
{\sl 5.1. The symmetric two-state chain.}
Consider a chain with transition matrix 
$$\mtrix{1-\theta &\theta \cr \theta &1-\theta \cr}. $$
The likelihood 
$\theta^{N_{0,1}+N_{1,0}}(1-\theta)^{N_{0,0}+N_{1,1}}$
is maximised for $\hatt\theta_\ML=(N_{0,1}+N_{1,0})/n$. 
One finds $u_{0,0}=-1/(1-\theta)=u_{1,1}$ 
while $u_{0,1}=1/\theta=u_{1,0}$, leading to 
$J=\{\theta(1-\theta)\}^{-1}$; in particular
the limit distribution of $\rootn(\hatt\theta_\ML-\theta)$
is $\N(0,\theta(1-\theta))$. 
This particular chain is symmetric with stationary distribution
$(\half,\half)$. By (3.1) the QL method therefore becomes
equivalent to the ML method. 

The PL method is however different. From 
$$P^2=\mtrix{(1-\theta)^2+\theta^2 &2\theta(1-\theta) \cr
   2\theta(1-\theta) &(1-\theta)^2+\theta^2 \cr} $$
one finds that the pseudo-likelihood becomes
$$\pl_n(\theta)=
 \Bigl({\theta^2\over \theta^2+(1-\theta)^2}\Bigr)^{N_{0,1,0}+N_{1,0,1}}
 \Bigl({(1-\theta)^2\over \theta^2
   +(1-\theta)^2}\Bigr)^{N_{0,0,0}+N_{1,1,1}},$$
which is maximal when 
$${\hatt\theta^2\over \hatt\theta^2+(1-\hatt\theta)^2}
  ={N_{0,1,0}+N_{1,0,1}\over N_{0,\cdot,0}+N_{1,\cdot,1}}=\rho_n
  \quad {\rm or} \quad 
  \hatt\theta={\rho_n^{1/2}\over \rho_n^{1/2}+(1-\rho_n)^{1/2}}. $$
Here $w_{0,0}$ and $w_{1,1}$ are equal to 
$(4\theta-2)/p_{0,0}^{(2)}$ while $w_{0,1}$ and $w_{1,0}$ 
are equal to $(-4\theta+2)/p_{0,1}^{(2)}$. This leads to 
$M=4(2\theta-1)^2\{1/p_{0,0}^{(2)}+1/p_{0,1}^{(2)}\}$.  
Also required for the limit variance 
of the PL estimator is $Q$ of (4.2), which with
some efforts is found to be
$$Q=\{(1-\theta)^3+\theta^3\}w^2_{0,0}
   +(1-\theta)\theta (2w_{0,0}w_{0,1}+w^2_{0,1}), $$
where 
$w_{0,0}=(4\theta-2)/p_{0,0}^{(2)}$, 
$w_{0,1}=(-4\theta+2)/p_{0,1}^{(2)}$, and 
$p_{i,j}^{(2)}$ denotes the $(i,j)$-th element 
of the two-step transition matrix $P^2$.
In fact one learns, from applying the PL proposition, 
that the PL estimator has a constant asymptotic variance
$1/4$; that is, $\sqrt{n}(\hatt\theta_\PL-\theta)$ 
converges in distribution to $\N(0, 1/4)$.

The asymptotic relative efficiency (ARE) of the PL is here 
$${\var(\ML)\over \var(\PL)}={1\over 4\theta(1-\theta)}, $$
showing that the PL loses rather significantly
in comparison with the ML when $\theta$ moves
towards values of strong spatial dependence,
i.e.~$\theta$ closer to 0 or 1. 
Simulation work, not reported on here, confirm
that distributions of both estimators reasonably 
quickly converge to normality and that the
large-sample approximations to standard deviations
are fairly accurate even for moderate $n$. 

\subsection
{\sl 5.2. The general two-state chain.}
Now let the transition matrix be 
$$P=\mtrix{1-\alpha &\alpha \cr \beta &1-\beta \cr}. \eqno(5.1)$$
Here 
$$P^k={1\over \alpha+\beta}\mtrix{\beta &\alpha \cr \beta &\alpha \cr}
   +{(1-\alpha-\beta)^k\over \alpha+\beta}
   \mtrix{\alpha &-\alpha \cr -\beta &\beta \cr}
   \quad {\rm for\ }k\ge0, $$
see e.g.~Karlin and Taylor (1975, p.~78); 
in particular, 
$p_0=\beta/(\alpha+\beta)$ and $p_1=\alpha/(\alpha+\beta)$.

We pause for a minute to remark that Markov chains first
saw light in Markov (1906), and that the first instance
ever of actual Markov chain modelling and data analysis
is Markov (1913). Astoundingly, he went through all
of the first 20,000 letters of Puskhin's {\sl Yevgeni{\u\i } Onegin},
and tabled transitions from vowels to vowels and consonants
and from consonants to vowels and consonants;
see (5.2). 

His results, in effect fitting model (5.1) to the 
Pushkin poem, were $1-\hatt\alpha=.128$ from vowel to vowel
and $\hatt\beta=.663$ from consonant to vowel,
leading also, he observed, to equilibrium frequencies
$.432$ for vowels and $.568$ for consonants. 
$${\rm Pushkin:}\,\,\,\matrix{1104 & 7534 \cr 
                        7533 & 3829 \cr}, \qquad
{\rm in\ English:}\,\,\,\matrix{1484  & 6396 \cr 
                          6397  & 5723 \cr}. \eqno(5.2)$$
{\smallskip\narrower\noindent\sl\baselineskiptables 
{\csc Table 5.0.}
Markov's data. 
\smallskip}


Markov's estimates were in our notation 
$\hatt\alpha=N_{01}/(N_{00}+N_{01})$
and $\hatt\beta=N_{10}/(N_{10}+N_{11})$. These may 
be recognised as the ML estimators, maximising
$$\log\lik_n=N_{00}\log(1-\alpha)+N_{01}\log\alpha
   +N_{10}\log\beta+N_{11}\log(1-\beta). $$
The QL estimators maximise 
$$\log\ql_n=(n-N_{00})\log\alpha+N_{00}\log(1-\alpha)
   +(n-N_{11})\log\beta+N_{11}\log(1-\beta)
   -n\log(\alpha+\beta). $$
Markov might have been surprised to see that 
the ML and QL estimation strategies, while different, 
are asymptotically equivalent. 

\centerline{\includegraphics[scale=0.39]{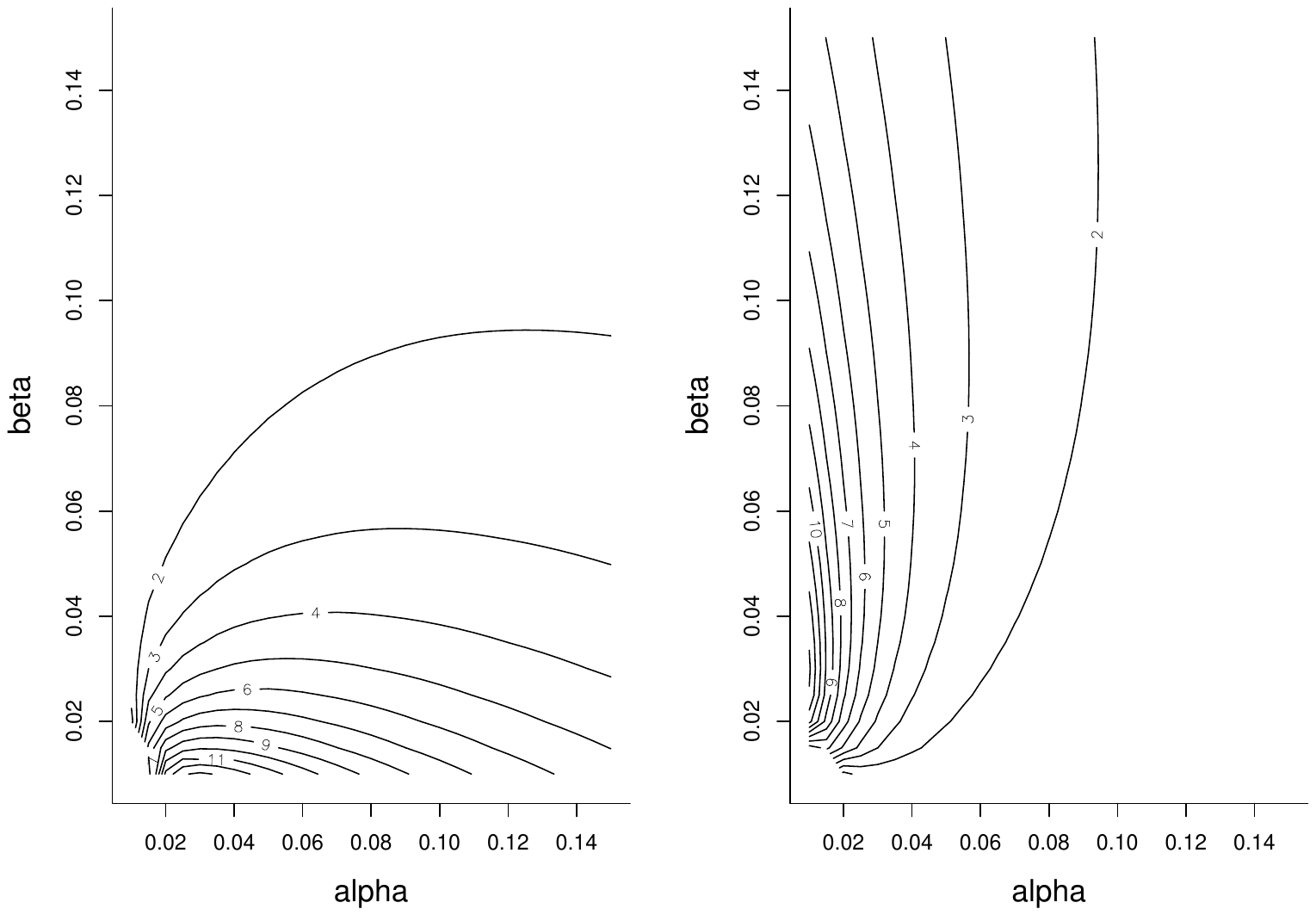}}

{\smallskip\narrower\sl\noindent\baselineskip11pt
{\csc Figure 5.1.}
Contour plots of the variance ratio for PL with respect to ML,
for estimation of $\alpha$ (left panel) and of $\beta$
(right panel), for the $[0,0.15]\times[0.15]$ subset
of the parameter space. 
\smallskip}

The quantities of (2.4) and (3.3) are found to be 
$$\eqalign{
\mtrix{\gamma_{0,0} &\gamma_{0,1} \cr 
      \gamma_{1,0} &\gamma_{1,1}}
&={1\over (\alpha+\beta)^2}\mtrix{\alpha &-\alpha \cr -\beta &\beta \cr}, \cr 
\mtrix{\bar\gamma_{0,0} &\bar\gamma_{0,1} \cr 
      \bar\gamma_{1,0} &\bar\gamma_{1,1}}
&={1-\alpha-\beta\over (\alpha+\beta)^2}
   \mtrix{\alpha &-\alpha \cr -\beta &\beta \cr}. \cr}$$
Furthermore, 
$$u_{0,0}=\mtrix{-1/(1-\alpha) \cr 0 \cr},\,\,
  u_{0,1}=\mtrix{1/\alpha \cr 0 \cr},\,\, 
  u_{1,0}=\mtrix{0 \cr 1/\beta \cr},\,\, 
  u_{1,1}=\mtrix{0 \cr -1/(1-\beta) \cr}, $$
with consequent formulae for $J_0$, $J_1$ and 
$$J=p_0J_0+p_1J_1
   =\mtrix{p_0/\{\alpha(1-\alpha)\}&0 \cr 0&p_1/\{\beta(1-\beta)\} \cr}. $$
This shows that $\hatt\alpha$ and $\hatt\beta$ 
are asymptotically independent, with limiting 
normalised standard deviations equal to 
$\alpha(1-\alpha)/p_0$ and $\beta(1-\beta)/p_1$. 


For the QL strategy, 
$$v_0=\mtrix{0 \cr 1/\beta \cr}-w
  \quad {\rm and} \quad 
  v_1=\mtrix{1/\alpha \cr 0 \cr}-w 
  \quad {\rm where} \quad 
  w={1\over \alpha+\beta}\mtrix{1 \cr 1 \cr}, $$
where some calculations give
$$G=p_0\bar\gamma_{0,0}v_0v_0^\tr
+p_0\bar\gamma_{0,1}v_0v_1^\tr
+p_1\bar\gamma_{1,0}v_1v_0^\tr
+p_1\bar\gamma_{1,1}v_1v_1^\tr
={1-\alpha-\beta\over (\alpha+\beta)^3}
   \mtrix{\beta/\alpha &-1 \cr -1 &\alpha/\beta \cr}. $$
One also finds
$$\eqalign{
H&=p_0v_0v_0^\tr+p_1v_1v_1^\tr
   ={1\over (\alpha+\beta)^2}\mtrix{\beta/\alpha & -1\cr 
                                    -1 &\alpha/\beta \cr}, \cr
L&=\sum_{a,b}p_ap_bu_{a,b}\kappa_b^\tr
  ={1\over (\alpha+\beta)^3}
  \mtrix{\beta(1+\beta/\alpha) &-(\alpha+\beta) \cr
      -(\alpha+\beta) &\alpha(1+\alpha/\beta) \cr}=H. }$$
One may now, with the required linear algebraic patience, 
calculate the asymptotic variance matrix from
the QL proposition, and verify that it is identical
to the $J^{-1}$ of the ML proposition. 

For the PL story, note that 
$$P^2=\mtrix{(1-\alpha)^2+\alpha\beta &\alpha(2-\alpha-\beta) \cr
   \beta(2-\alpha-\beta) &\alpha\beta+(1-\beta)^2 \cr}. $$
Thus 
$$\eqalign{
w_{0,0}&=\mtrix{2\alpha-2+\beta \cr \alpha \cr}/p_{0,0}^{(2)}, \quad
  w_{0,1}=\mtrix{-2\alpha+2-\beta \cr -\alpha \cr}/p_{0,1}^{(2)}, \cr
w_{1,0}&=\mtrix{-\beta \cr -2\beta-\alpha+2 \cr}/p_{1,0}^{(2)}, \quad
  w_{1,1}=\mtrix{\beta \cr \alpha+2\beta-2 \cr}/p_{1,1}^{(2)}. \cr}$$ 
This makes it possible to calculate the matrices 
$M$ and $Q$ featuring in the PL Proposition. 
We have made computer programmes that find 
and compare the limit variance matrices for the PL and the ML. 
The PL and ML (and the QL) are large-sample equivalent 
when $\alpha+\beta=1$, and the variance increase of 
PL is not large in the vicinity of this strip. When 
$(\alpha,\beta)$ turn towards stronger spatial dependence,
however, the variance ratios become noticeably bigger;
see Figure 5.1. We also carried out simulation experiments,
recording simulation mean and standard deviation for
all three simulation procedures, and found good agreement
with the asymptotics results. 
 
\subsection       
{\sl 5.3. An equicorrelation chain.} 
Consider a transition matrix of the form
$$P=(1-\rho)p+\rho I, $$
where $p$ is a probability vector;
thus $p_{a,b}=(1-\rho)p_b + \rho\delta_{a,b}$. 
Here one finds $P^k=(1-\rho^k)p+\rho^kI$,
so that $p$ indeed is the stationary distribution of the 
chain, consistent with our $p_a$ notation. One finds
$$\gamma_{a,b}={1\over 1-\rho}(\delta_{a,b}-p_b)
  \quad {\rm and} \quad
  \bar\gamma_{a,b}={\rho\over 1-\rho}(\delta_{a,b}-p_b). $$
One may now consider estimation of $\rho$ and 
the $p_a$s, with or without structure 
inside the $p_a$ distribution. 

\centerline{\includegraphics[scale=0.39]{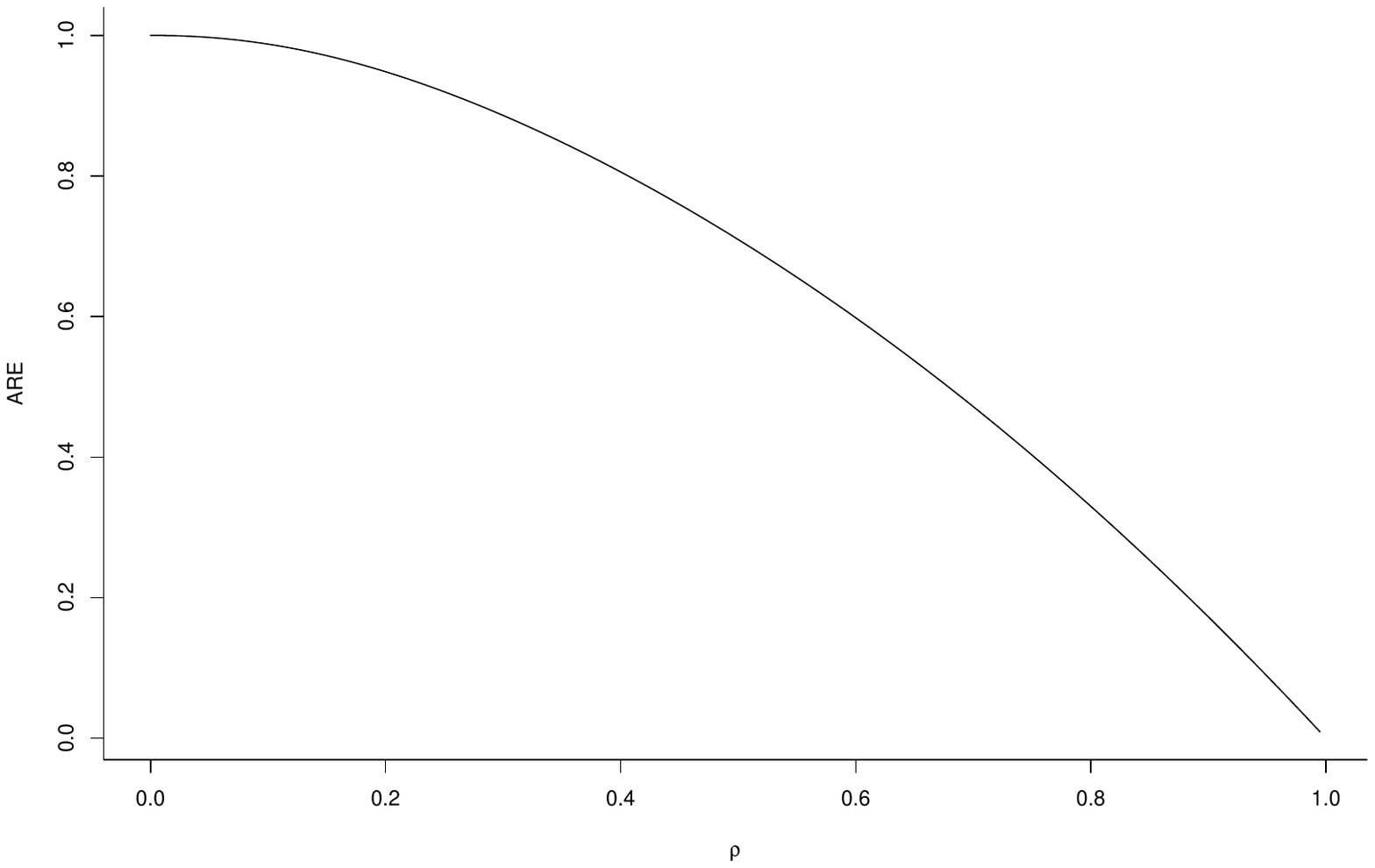}}

{\smallskip\narrower\sl\noindent\baselineskip11pt
{\csc Figure 5.2.}
Asymptotic relative efficiency for the PL method,
with respect to the ML and QL method, 
for a three-stage equicorrelation chain, 
as a function of~$\rho$.  
\smallskip} 

We start by assuming $p=(p_1,\ldots,p_k)^\tr$ known 
and consider estimation of the parameter $\rho$. 
We have $u_{a,b}=(\delta_{a,b}-p_b)/p_{a,b}(\rho)$ and
$$ J=\sum_{a,b}  p_a { (\delta_{a,b}-p_b)^2 \over p_{a,b}(\rho)}=
\sum_{a,b} p_a{ (\delta_{a,b}-p_b)^2 \over (1-\rho)p_b+\rho\delta_{a,b}}.$$
Since the stationary distribution $p$ is known, 
it does not depend on $\rho$ and we have $v_a=0$ 
for $a=1,\ldots,k$. In consequence, the QL is fully 
equivalent to ML. Then consider the PL method. We have 
$w_{a,b}=2\rho(\delta_{a,b}-p_b)/p^{(2)}_{a,b}(\rho)$, 
where $p^{(2)}_{a,b}(\rho)=(1-\rho^2)p_b+\rho^2\delta_{a,b}$. 
The relevant matrices are found to be 
$$\eqalign
{M&=4\rho^2 \sum_{a,c} 
p_a {(\delta_{a,c}-p_c)^2 \over p^{(2)}_{a,c}(\rho)}, \cr
Q&=4\rho^2 \sum_{a,c,d,f} p_a p_{a,d}(\rho) 
  p_{d,c}(\rho) p_{c,f}(\rho) 
  {\delta_{a,c}-p_c\over p^{(2)}_{a,c}(\rho)} 
  {\delta_{d,f}-p_f\over p^{(2)}_{d,f}(\rho)}.}$$
For illustration, we take case of a three states 
chain with $p=(0.3, 0.6, 0.1)^\tr$ and different 
values for $\rho$. As perhaps expected, the two methods are 
large-sample equivalent for $\rho=0$, while 
the efficiency of the PL strategy (the variance ratio
of the ML estimator to the PL estimator) decays 
exponentially to zero when the value of $\rho$
is increased, see Figure 5.2.

When the starting probabilities $p$ are also unknown
and part of the estimation task,  
QL is not anymore equivalent to ML. 
Its efficiency loss is very modest, however, 
as seen in this illustration, featuring 
standard deviations of the limit distributions
for the three different methods. Again the 
PL methods does poorly, see Table 5.1.
$$\matrix{
     &\ML    &\QL    &\PL \cr
\rho &0.676  &0.678  &1.011 \cr
p_1  &0.583  &0.607  &1.178 \cr
p_2  &0.693  &0.708  &1.367 \cr} $$
{\smallskip\narrower\noindent\sl\baselineskiptables 
{\csc Table 5.1.}
For parameters $\rho=0.5,p_1=0.3,p_2=0.6,p_3=0.1$, 
the table displays the limit distribution standard deviation 
for the ML, QL and PL methods.
\smallskip}

\subsection
{\sl 5.4. A three-state example.}
Consider a Markov chain with transition probability matrix
$$\mtrix{1-\alpha-\beta &\alpha &\beta \cr
           \alpha &1-\alpha-\beta &\beta \cr
           \alpha &\beta &1-\alpha-\beta \cr}, $$
which with some efforts is seen to have equilibrium distribution 
$${\alpha\over 2\alpha+\beta},\quad 
  {\alpha^2+\alpha\beta+\beta^2\over (\alpha+2\beta)(2\alpha+\beta)},\quad 
  {\beta\over \alpha+2\beta}. $$
We find the explicit formula 
$$J=
\pmatrix{
1/(1-\alpha-\beta)+1/\alpha & 1/(1-\alpha-\beta) \cr
1/(1-\alpha-\beta) & 1/(1-\alpha-\beta)+1/\beta \cr}, $$
while the other matrices $G$, $H$, $L$ for the QL method
and $M$, $Q$ for the PL method can be computed numerically,
with some programming efforts. 
In particular, to evaluate the $k$-steps transition 
probability matrix $P^k$, one can use 
an eigenvalue decomposition. 
As an illustration, take $(\alpha,\beta)$ 
equal to $(0.21,0.55)$. Then the ML standard deviations
are $0.407$ and $0.497$
and the QL standard deviations $0.431$ and $0.517$,
while the PL standard deviations are $0.487$ and $0.567$, 
signalling efficiency loss. 
Simulating 1000 chains of length $n=500$ from the model, 
at this parameter point, gave means and standard deviations
in good agreement with the large-sample theory. 

\subsection
{\sl 5.5. The one-dimensional Ising model.} 
The next example is a binary one-dimensional version 
of the Ising model; for the two-dimensional,
which goes back to 1925, see e.g.\allowbreak~Pickard (1987). 
The model is defined in terms of the conditional probabilities
$$\Pr\{(X_i=x_i\midd X_{i-1}=x_{i-1}, X_{i+1}=x_{i+1}\}
   \propto\exp\Bigl[\beta\!\!\sum_{j\in\{i-1,i+1\}}\!\! 
  I\{x_j=x_i\}\Bigr] \quad {\rm for\ }x_i \in \{0,1\}, $$
for $i=1,\ldots,n-1$. This corresponds to the transition matrix 
$$P={1\over 1+\exp(\beta)}
\pmatrix{\exp(\beta) & 1 \cr
         1 & \exp(\beta) \cr},$$
while the two-steps transition matrix is given by
$$P^2= {1 \over\{1+\exp(\beta)\}^2}
\pmatrix{1+\exp(2\beta) & 2\exp(\beta) \cr
         2\exp(\beta) & 1+\exp(2\beta) \cr}. $$
The transition matrix being symmetric, 
the QL method coincides with the ML. One finds 
$$\hatt\beta_\ML=\log{N_{0,0}+N_{1,1} 
   \over N_{0,1}+N_{1,0}}
  \quad {\rm and} \quad 
  \hatt\beta_\PL=\half
  \log{N_{0,0,0}+N_{1,1,1}\over N_{0,1,0}+N_{1,0,1}}. $$
Furthermore, 
$$J={\exp(\beta)\over \{1+\exp(\beta)\}^2}
  \quad {\rm and} \quad  
  M={2\exp(\beta)\{1-\exp(\beta)\}^2
    \over \{1+\exp(\beta)\}^2\{1+\exp(2\beta)\}}, $$
while the $Q$ matrix of (4.2) has been computed 
numerically in our illustrations. Some values
of the limiting standard deviations 
for the ML (and hence QL) and the PL are as illustrated
in the Table 5.2.
$$\matrix{
\beta     &\ML    &\PL \cr
0.0       &2.000  & 2.000 \cr       
0.5       &2.063  & 2.128 \cr
1.0       &2.255  & 2.543 \cr
1.5       &2.589  & 3.352 \cr
3.0       &4.705  &11.068 \cr}$$
{\smallskip\narrower\noindent\sl\baselineskiptables 
{\csc Table 5.2.}
For different values of $\beta$,
the table displays the limit
distribution standard deviation for the 
ML and PL methods. 
\smallskip}

As seen for the other cases, the PL method loses 
increasingly as the dependency strengthens.
Various simulation exercises, not reported 
on in detail here, supported the large-sample
theory, and indicated reasonably rapid 
convergence to normality for both methods. 

\subsection
{\sl 5.6. A random walk with two reflecting barriers.} 
Let us consider the Markov chain with transition matrix
$$ P=\pmatrix{
 0 & 1 & 0 & 0 & \cdots & 0 \cr
q & 0 & p & 0 & \cdots & 0 \cr
0 & q & 0 & p & \cdots & 0\cr
\vdots & \vdots & \vdots & \vdots & \ddots& \vdots\cr
0 & \cdots & \cdots & 0 & 1 & 0\cr}, $$
where $p+q=1$. It has two reflecting barriers 
and say $k$ states. From $\pi P=\pi$
and some work it follows that the stationary distribution 
is given by
$$\eqalign{
\pi_1 &=\{1+S_{k-2}/p+(p/q)^{k-2}\}^{-1}, \cr
\pi_i &= \pi_1p^{i-2}/q^{i-1} \quad {\rm for\ }i=2,\ldots,k-1, \cr
\pi_k &= \pi_1 (p/q)^{k-2}, \cr}$$
where $S_{k-2}=\sum_{i=1}^{k-2} (p/q)^i.$
The two-steps transition matrix is easily found. 
The log-likelihood function is
$$
\log\lik_n(p)=\sum_{i=2}^{k-1} 
  N_{i,i+1}\log p+\sum_{i=2}^{k-1} N_{i,i-1}\log(1-p), $$
which is maximal for 
$$\hatt p_\ML={\sum_{i=2}^{k-1} N_{i,i+1}
   \over \sum_{i=2}^{k-1} (N_{i,i-1}+N_{i,i+1})}.$$
Moreover, we find $J=(1-\pi_1-\pi_k)/(pq)$, 
which means that the limit distribution for 
the ML method has variance $pq/(1-\pi_1-\pi_k)$. 


\centerline{\includegraphics[scale=0.39]{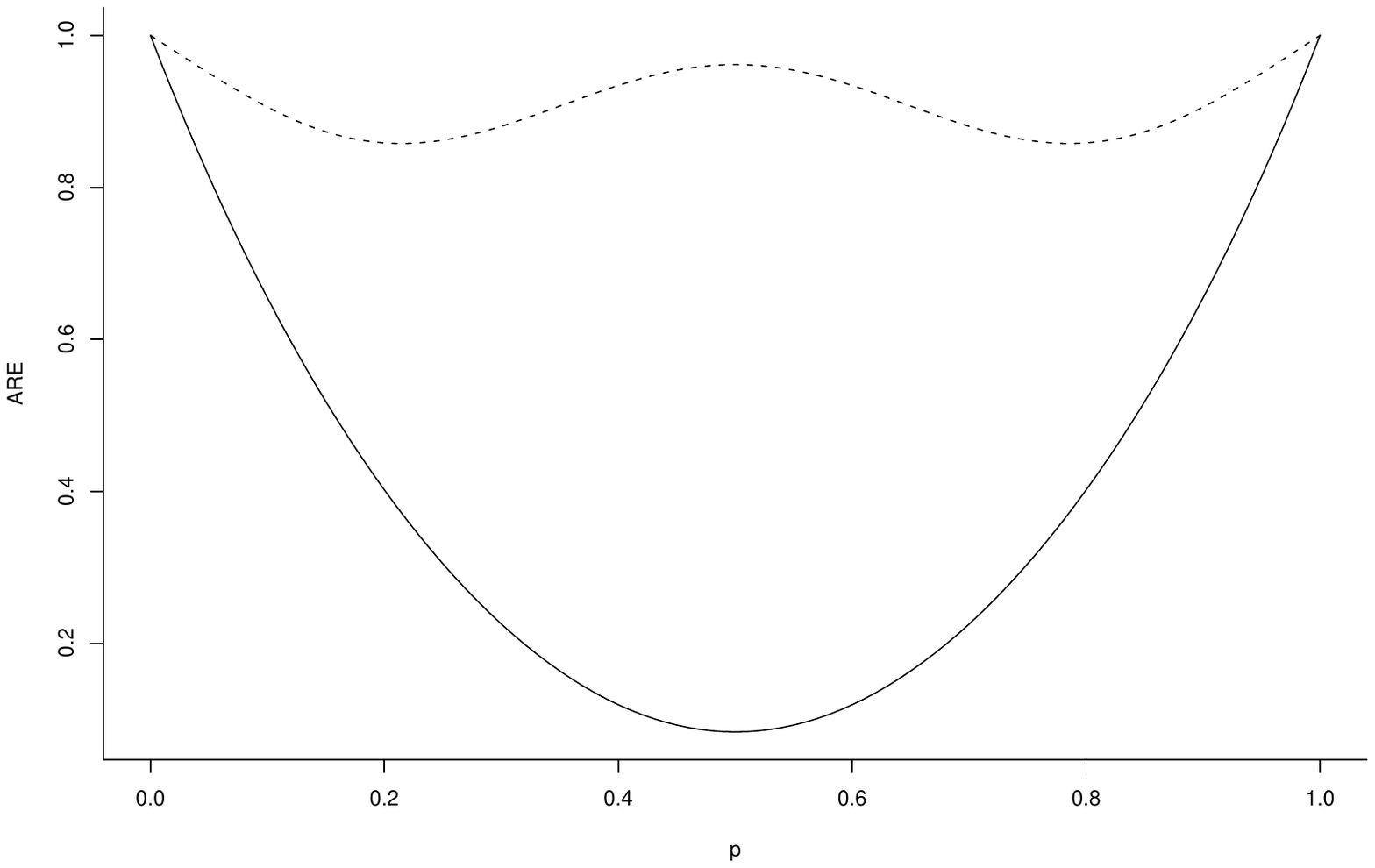}}

{\smallskip\noindent\sl\baselineskiptables
{\csc Figure 5.3.} 
The random walk with two reflecting barriers: six states example.
The solid line correspond to the ARE for PL, 
while the dashed one to QL.
\smallskip}


\centerline{\includegraphics[scale=0.39]{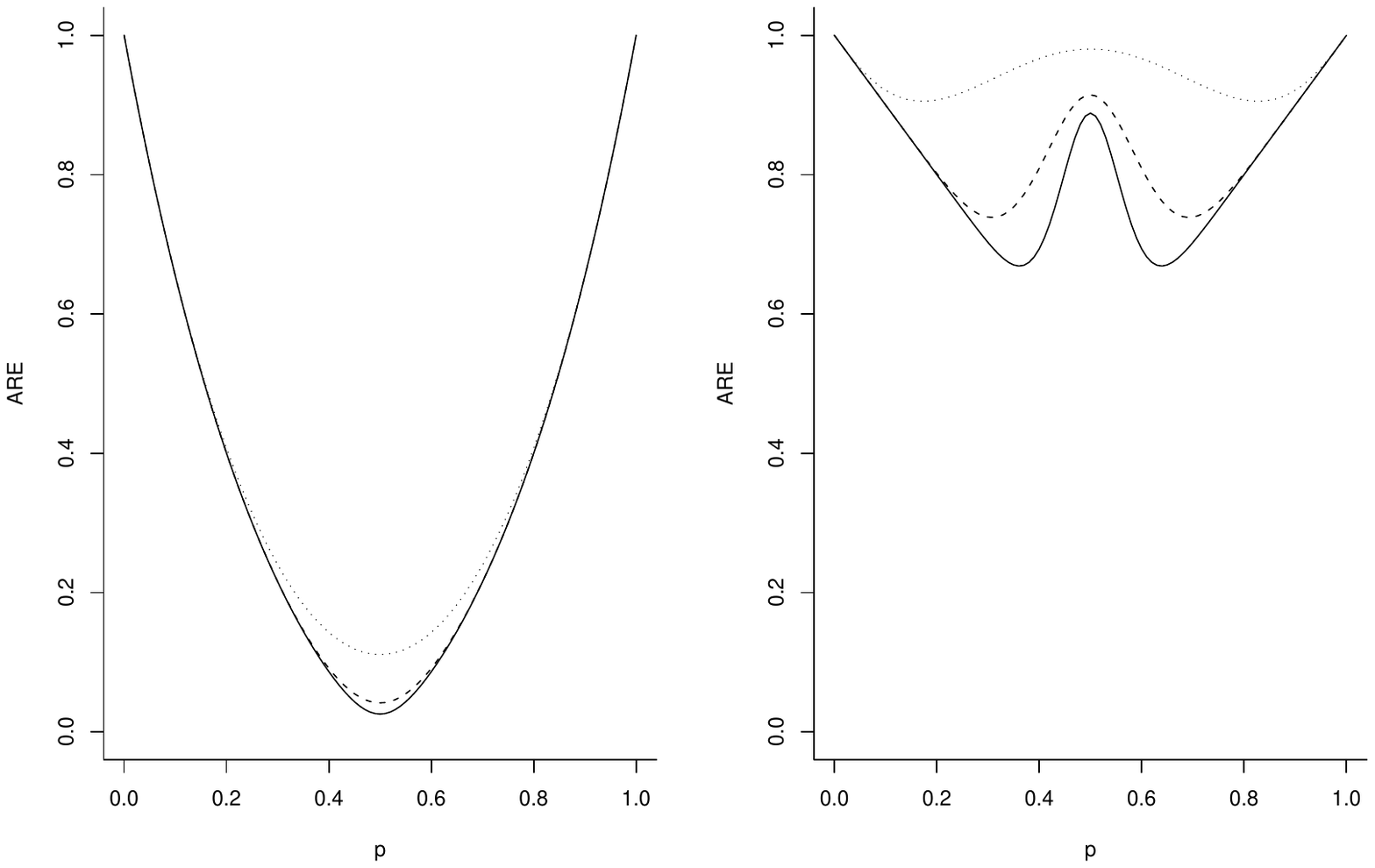}}

{\smallskip\noindent\sl\baselineskiptables
{\csc Figure 5.4.} 
The random walk with two reflecting barriers: 
the effects of increasing the number of states on the 
PL estimator of $p$ [left panel] 
and on the QL estimator of $p$ [right panel]. 
The solid line corresponds to a 15 states chain; 
the dashed line to 10 states; and 
the dotted line to 5 states.
\smallskip}

In order to compute the matrices involved 
in the QL and PL computations, note that 
$$\eqalign{
v_1&=  -\pi_1\Bigl\{ 
{1\over pq}\sum_{i=1}^{k-2}\Bigl({p\over q} \Bigr)^{i-1}
\Bigl({i \over q}-1\Bigr)   
+(k-2)\Bigl({p\over q}\Bigr)^{k-3}{1 \over q^2}\Bigr\},   \cr
v_i &=  v_1-{1\over p}+(i-1){1 \over pq} 
   \quad {\rm for\ }i=2,\ldots,k-1,\cr
v_k &= v_1+(k-2){1\over pq}. \cr}$$
Furthermore, the matrices containing the $w_{a,b}$s 
and $u_{a,b}$s are respectively 
$$
\pmatrix{
-1/q & 0 & 1/p & 0 & 0 & 0 & \cdots & 0\cr
0 & -2p/(q+qp) & 0 & 2/p & 0 & 0  & \cdots & 0\cr
-2/q & 0 & (1-2p)/(qp) & 0 & 2/p & 0 & \cdots & 0\cr
0 & -2/q & 0 & (1-2p)/(qp) & 0 & 2/p & \cdots & 0\cr
\vdots & \vdots & \vdots & \vdots & \vdots & \vdots & \ddots & \vdots\cr
0 & \cdots & \cdots & \cdots & -2/q & 0 & {2(1-p)\over qp+p} & 0 \cr
0 & \cdots & \cdots & \cdots & 0 & -1/q & 0 & 1/p \cr } $$
and 
$$\pmatrix{
 0 & 0 & 0 & 0 & \cdots & 0 \cr
-1/q & 0 & 1/p & 0 & \cdots & 0 \cr
0 & -1/q & 0 & 1/p & \cdots & 0\cr
\vdots & \vdots & \vdots & \vdots & \ddots& \vdots\cr
0 & \cdots & \cdots & 0 & 0 & 0\cr}. $$
The $\gamma_{a,b}$ and $\bar{\gamma}_{a,b}$ values
were computed using the eigenvalues decomposition. 

\centerline{\includegraphics[scale=0.39]{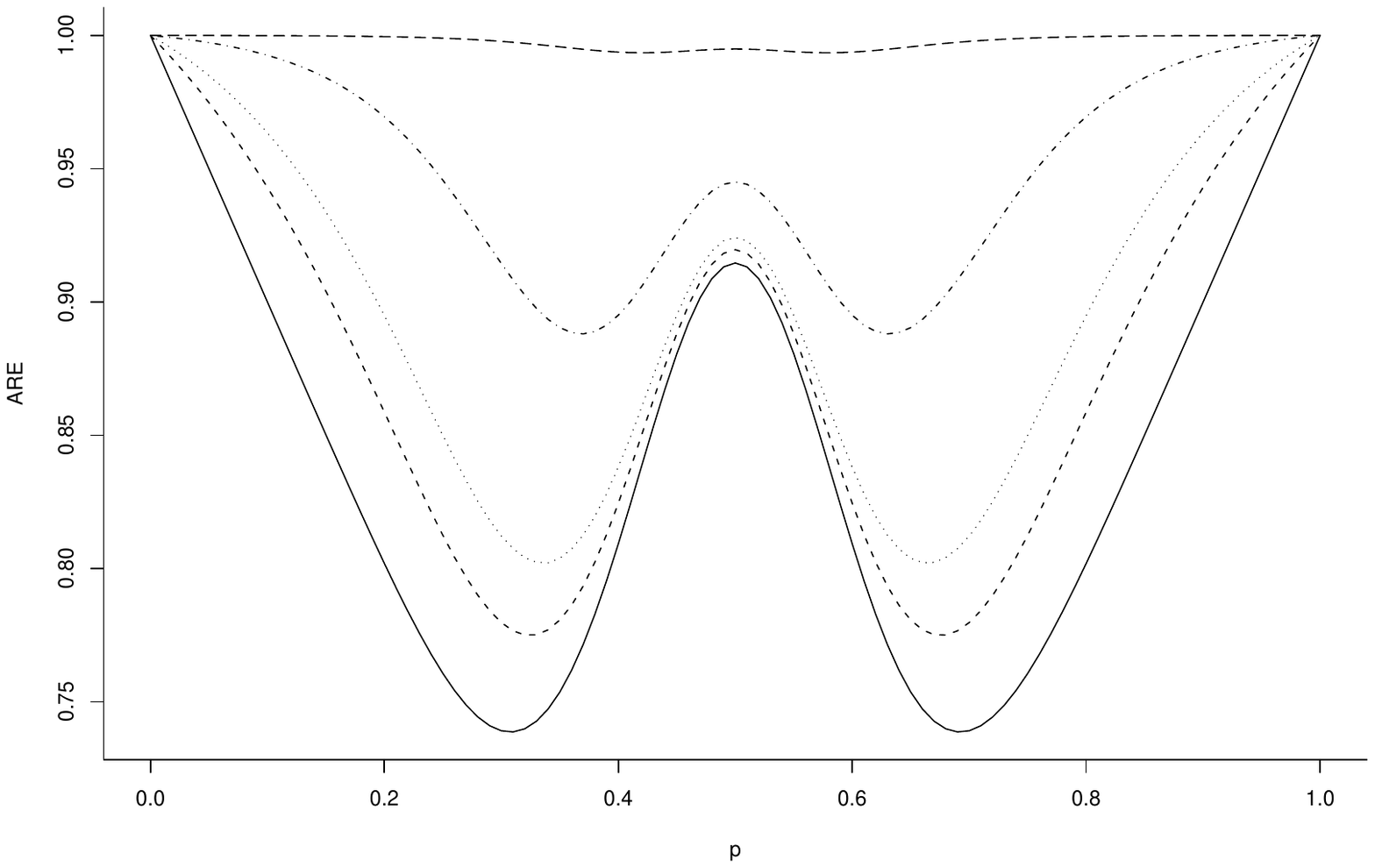}}

{\smallskip\noindent\sl\baselineskiptables
{\csc Figure 5.5.} 
The random walk two reflecting barriers: ten states chain.
The curves correspond to the ARE for QL 
of order 2, 3, 4, 10 and 100 (growing).
\smallskip}

Again, the matrices involved in describing 
the large-sample behaviour of the QL and PL
methods may now be computed numerically. 
To illustrate different aspects involved
in the comparison, we varied the $p$ parameter
as well as the number $k$ of states, and 
examined the large-sample standard deviations
of the methods, as well as the usual 
asymptotic relative efficiency ARE.
Figure 5.3 shows ARE as a function of $p$
for a situation with six states, with QL close
to ML and PL doing poorly. 
Figure 5.4 (left panel) shows that the ARE curve for PL
has basically the same shape and almost the same
values, for different number of states. 
The right panel shows firstly (once again) 
that the QL does much better than the PL, 
but also that it loses somewhat in efficiency 
when the number of states grow. 
This invites trying out QL estimation of higher order. 
The effect is displayed in Figure 5.5. 
Also for these chains have we carried out simulations
to test adequacy of the large-sample predictions,
and again there is basic agreement, for 
chain lengths $n=100$ and more. 
\eject



\section
\centerline{\bf 6. Markov chain models for DNA sequence evolution}

\hop
Markov chains are fruitful models for describing the evolution of DNA
sequences. 
The states of the chain in question are the four DNA nucleotides 
adenine (A), guanine (G), cytosine (C), thymine (T). 
These may be grouped into `type 1', the 
purine nucleotides A and G, and `type 2', 
the pyrimidine nucleotides C and T. 
DNA evolutionary models are considered for studying
the time dynamics of two homologous sequences 
hypothesised to having come from a common ancestor. 
With {\it substitutions} is meant changes occurring
between two nucleotides located at the same site
within two sequences. A substitution that preserves 
the type is called a {\it transition}, otherwise 
it is called a {\it transversion}. 
In DNA, transitions are much more common than transversions. 

\subsection
{\sl 6.1. Kimura models.}
In the following we choose to illustrate our general
methods and results for the four-parameter model 
with transition matrix
$$\matrix{
\niente&A      &G      &C     &T \cr
A      &1-2\alpha-\gamma &\gamma      &\alpha      &\alpha \cr
G      &\delta      &1-2\alpha-\delta &\alpha      &\alpha \cr 
C      &\beta      &\beta      &1-2\beta-\gamma &\gamma \cr
T      &\beta      &\beta      &\delta      &1-2\beta-\delta \cr} 
   \eqno(6.1)$$
in terms of positive parameters $\alpha,\beta,\gamma,\delta$ 
such that $2\alpha+\gamma<1$, $2\alpha+\delta<1$, 
$2\beta+\gamma<1$, $2\beta+\delta<1$. 
There are several special cases and generalisations 
worth mentioning. The simplest such model is the 
Jukes--Cantor one which has $\alpha=\beta=\gamma=\delta$.
Then there are several so-called Kimura models, 
where the basic one has two parameters, distinguishing
between transitions and transversions; 
see e.g.~Kimura (1980, 1981), Durrett (2002), 
and Hobolth and Jensen (2005) for some further discussion of models. 
Version (6.1), which we shall work with here, corresponds to 
the so-called Blaisdell model; see Blaisdell (1985). 
Its equilibrium distribution can be found with 
some algebraic efforts, starting from the equations $p P=p$; 
$$\eqalign{
p_A&={\beta\over \alpha+\beta}{\alpha+\delta\over
  2\alpha+\gamma+\delta}, 
  \qquad
  p_G={\beta\over \alpha+\beta}{\alpha+\gamma\over 2\alpha+\gamma+\delta}, \cr
p_C&={\alpha\over \alpha+\beta}{\beta+\delta\over
  2\beta+\gamma+\delta},
  \qquad 
  p_T={\alpha\over \alpha+\beta}{\beta+\gamma\over
  2\beta+\gamma+\delta}. \cr}\eqno(6.2)$$
DNA sequence evolution studies are complicated by the fact 
that direct observation of the nucleotides transitions is 
unavailable. Instead, the transitions rates are inferred
from the pairwise comparison of two homologous DNA sequences, 
see e.g.~Hobolth and Jensen (2005) and references therein. 
DNA sequence data that more properly may be seen as having arisen 
as direct transition counts can also be found in the literature, 
see e.g.~Davison (2003, Ch.~6). Interestingly, Davison's analysis 
(which focusses on aspects different from those considered here) 
supports the one-step memory Markov model. 

We have investigated the ML, PL, QL performance for 
the (6.1) model via simulations at various positions 
in the parameter space, and choose for the purposes
of the current report to present results corresponding 
to a fixed set of `typical values' of $(\alpha,\beta,\gamma,\delta)$,
namely $(.027,.041,.123,.128)$. These values stem from
analysis of a real data set. Performance statements 
very similar to those arrived at below were reached
also for various other parameter combinations.  

The ML estimators are found by maximising 
$$\eqalign{
\log\lik_n(\theta)
&=N_{AA}\log(1-2\alpha-\gamma)+N_{GG}\log(1-2\alpha-\delta)
  +N_{CC}\log(1-2\beta-\gamma) \cr 
&\qquad +N_{TT}\log(1-2\beta-\delta)
  +N_{1,2}\log \alpha+N_{2,1}\log \beta \cr
&\qquad 
  +(N_{AG}+N_{CT})\log \gamma
  +(N_{GD}+N_{TC})\log \delta, \cr}\eqno(6.3)$$
where $N_{1,2}=N_{AC}+N_{AT}+N_{GC}+N_{GT}$
is the number of transitions from type 1 to type 2,
and $N_{2,1}=N_{CA}+N_{CG}+N_{TA}+N_{TG}$
the number of transitions from type 2 to type 1;
also, we use $\theta$ to indicate the parameter
vector $(\alpha,\beta,\gamma,\delta)$. 
The QL estimators correspondingly 
maximise 
$$\log\ql_n(\theta)=\log\lik_n(\theta)
  +N_A\log p_A(\theta)+N_G\log p_C(\theta)
  +N_C\log p_C(\theta)+N_T\log p_T(\theta). \eqno(6.4)$$
 
The precision of the ML estimators is determined by 
$J^{-1}$, where 
$J=p_AJ_A+p_GJ_G+p_CJ_C+p_TJ_T$, and where 
the (2.2) recipe gives
$$\eqalign{
J_A&=\diag\Bigl({4\over 1-2\alpha-\gamma}+{2\over\alpha},0,
   {1\over 1-2\alpha-\gamma}+{1\over \gamma},0\Bigr)
   +{2\over 1-2\alpha-\gamma}e_{1,3}, \cr
J_G&=\diag\Bigl({4\over 1-2\alpha-\delta}+{2\over\alpha},0,
   0,{1\over 1-2\alpha-\delta}+{1\over \delta}\Bigr)
   +{2\over 1-2\alpha-\delta}e_{1,4}, \cr
J_C&=\diag\Bigl(0,{4\over 1-2\beta-\gamma}+{2\over\beta},
   {1\over 1-2\beta-\gamma}+{1\over \gamma},0\Bigr)
   +{2\over 1-2\beta-\gamma}e_{2,3}, \cr
J_T&=\diag\Bigl(0,{4\over 1-2\beta-\delta}+{2\over\beta},0,
   {1\over 1-2\beta-\delta}+{1\over \delta}\Bigr)
   +{2\over 1-2\beta-\delta}e_{2,4}, \cr}$$
in which $e_{i,j}$ is the $4\times4$ matrix 
with all zeroes apart from value 1 at positions 
$(i,j)$ and $(j,i)$. The precision of QL and PL 
estimators are similarly described by the propositions
of Sections 3 and 4, engaging matrices $G$, $H$, $L$
and $M$, $Q$. We have found these numerically
via programming, and compared large-sample precision 
of the three procedures in different regions of 
the parameter space. Again the picture emerges that 
(i) the PL often does poorly, 
(ii) the QL does almost as well as the ML. 
This is illustrated in Table 6.1, 
which shows limit distribution standard deviations
for the three methods, for each of the four parameters,
at the announced typical position $(.027,.041,.123,.128)$ 
in the parameter space. The table also includes
empirical means and $\rootn$ times standard deviations
for 1000 simulated chains. 
\def\hnow{\hskip12pt}
$$\matrix{
&      &\ML\colon   &     &        &\QL\colon   
       &      &       &\PL\colon        \cr
       &\true \hnow&\ave   &\sd   &\theory
       &\ave    &\sd   &\theory &\ave   &\sd   &\theory\cr
\alpha &.027 \hnow   &.028  &.158   &.146    
               &.028  &.156   &.146   &.031  &.615   &.622 \cr
\beta  &.041 \hnow &.042  &.232   &.215    
               &.042  &.234   &.216   &.048  &.769  &.931 \cr
\gamma &.122 \hnow &.123  &.441   &.458    
               &.122  &.453   &.475   &.117  &.779  &.878\cr
\delta &.126 \hnow &.128  &.475   &.472    
               &.129  &.493   &.489   &.126  &.837  &.907 \cr }$$

{\smallskip\narrower\noindent\sl\baselineskiptables 
{\csc Table 6.1.}
For parameters $(\alpha,\beta,\gamma,\delta)$ as 
indicated, the table displays empirical mean 
and $\sqrt{n}$ times empirical standard deviation
for 1000 simulated estimates, with the ML, QL, PL
estimators, for $n=500$. Also given is the limit
distribution standard deviation for the 
different methods. 
\smallskip}

\subsection
{\sl  6.2. Inference for other parameters.} 
For the Kimura type models each model parameter 
has a natural interpretation, and each of them
warrants scrutiny, e.g.~in the form of a confidence
interval. There are other parameters that can
be even more important, however, depending on
the focus of the study. The theory we have developed
for ML, QL, PL easily gives limit distribution
results also for such estimators, say 
$\hatt\psi=\psi(\hatt\theta)$ for some focus 
estimand $\psi=\psi(\theta)$, for each of 
the three estimation methods. This is simply 
the delta method: 
$$\eqalign{
\rootn(\hatt\psi_\ML-\psi)&\arr_d\N(0,\tau_\ML^2), \cr
\rootn(\hatt\psi_\QL-\psi)&\arr_d\N(0,\tau_\QL^2), \cr
\rootn(\hatt\psi_\PL-\psi)&\arr_d\N(0,\tau_\PL^2), \cr} 
   \eqno(6.5)$$
in which 
$$
\tau_\ML^2=\bar\psi^\tr\Sigma_\ML\bar\psi, \quad
\tau_\QL^2=\bar\psi^\tr\Sigma_\QL\bar\psi, \quad
\tau_\PL^2=\bar\psi^\tr\Sigma_\PL\bar\psi, $$
where $\bar\psi=\dell\psi(\theta)/\dell\theta$, 
and where $\Sigma$ is the limit distribution 
variance matrix for the method indicated by the subscript. 

Among several different quantities of interest
here, when analysing DNA data, are 
the probabilities of two consecutive pairs 
both being of type 1 (A or G), 
or both being of type 2 (C or T),
or experiencing a shift from type 1 to type 2,
or from type 2 to type 1. These probabilities 
are respectively  
$$p(1)=\beta(1-2\alpha)/(\alpha+\beta), \quad
  p(2)=\alpha(1-2\beta)/(\alpha+\beta), $$
$$p(1\arr2)=2\alpha\beta/(\alpha+\beta), \quad
  p(2\arr1)=2\alpha\beta/(\alpha+\beta). $$
Similarly one is sometimes interested in 
the ratio $\gamma/\delta$, for example for 
testing whether the rates involved might
be equal, or the mean time to transition from type 1 (A or G) 
to type 2 (C or T), and vice versa; 
these are found to be 
$$\psi_{1,2}={2\beta(\alpha+\beta)\over \alpha(1+2\beta)^2}
  \quad {\rm and} \quad 
  \psi_{2,1}={2\alpha(\alpha+\beta)\over \beta(1+2\alpha)^2}. $$
Each of these focus parameters can be estimated 
via the different methods, and the standard deviation 
estimated via (6.5). 
In view of the findings of Table 6.1, the QL based
confidence intervals would be about as short as 
those for the ML, while those based on the PL 
would be wider. 

Let us illustrate this with one such focus parameter,
namely the asynchronous distance between homologous
DNA sequences introduced in Barry and Hartigan (1987),
$\Delta=-(1/4)\log|P|$. It is shown there to estimate
the number of changes on the evolutionary path
from sequence to sequence, with each change weighted
by the inverse probability of the nucleotide changed from. 
For one of the simulated data sets considered in the previous 
section, using the ML method with the (6.1) model gives
$$\hatt\Delta_\ML
   =-(1/4)\log|P(\hatt\theta_\ML)|=0.2260. \eqno(6.6)$$
Its large-sample behaviour can be found from the above. From 
the matrix derivative fact that 
$\dell\log|P|/\dell P_{i,j}=P^{j,i}$, the $(j,i)$th
element of $P^{-1}$, follows 
$$\rootn\{\log|P(\hatt\theta)|-\log|P(\theta)|\}
\arr_d\sum_{i,j}P^{j,i}(\theta)V_{i,j}=\Tr\{P(\theta)^\tr V\}, $$
in terms of the limit versions $V_{i,j}$ 
of $\rootn\{P_{i,j}(\hatt\theta)-P_{i,j}(\theta)\}$.
In view of the structure of the Kimura model (6.1), 
the matrix of $V_{i,j}$ is seen to be 
$$V=\pmatrix{-2Y_1-Y_3 &Y_3 &Y_1 &Y_1 \cr
              Y_4 &-2Y_1-Y_4 &Y_1 &Y_1 \cr
              Y_2 &Y_2 &-2Y_2-Y_3 &Y_3 \cr
              Y_2 &Y_2 &Y_4 &-2Y_2-Y_4 \cr}, $$
where 
$$\rootn\pmatrix{\hatt\alpha-\alpha \cr \hatt\beta-\beta \cr
   \hatt\gamma-\gamma \cr \hatt\delta-\delta \cr}
   \arr_d\pmatrix{Y_1 \cr Y_2 \cr Y_3 \cr Y_4 \cr}. $$
This leads to 
$$\rootn\{\hatt\Delta_\ML-\Delta(\theta)\}
   \arr_d-(1/4)(c_1Y_1+c_2Y_2+c_3Y_3+c_4Y_4), $$
where 
$$\eqalign{
c_1&=-2P^{1,1}+P^{3,1}+P^{4,1}-2P^{2,2}+P^{3,2}+P^{4,2}, \cr
c_2&=P^{1,3}+P^{2,3}-2P^{3,3}+P^{1,4}+P^{2,4}-2P^{4,4}, \cr
c_3&=P^{2,1}-P^{1,1}-P^{3,3}+P^{4,3}, \cr
c_4&=P^{1,2}-P^{2,2}+P^{3,4}-P^{4,4}. \cr}$$
Applying these formulae one finds that the estimated
standard deviation of the limit distribution is 
$(1/4)\hatt c^\tr\hatt\Sigma\hatt c=0.563$, i.e.~standard 
error for the (6.6) estimate equal to 
$0.563/\rootn=0.0252$. This would be the basis of 
confidence intervals for Barry and Hartigan's 
asynchronous $\Delta$ distance. 
We carried out this analysis also with the QL method,
and got estimates and precision estimates very
similar to those using ML. 

The method we have used here, to illustrate our general
theory, is different from the one used by Barry and Hartigan,
since they did not assume any model for the 16 transition
probabilities. Using their in this sense nonparametric
approach leads to the estimate 0.2270 for $\Delta$, 
for the simulated data set mentioned above, 
and an estimate of limit distribution standard deviation
equal to 0.571, slightly bigger than with our model-based method.  

\smallskip
{\csc Remark.}
The (6.1) model should find use in many other applications,
where the basic feature is that of two main categories,
say I and II, that are further split into finer 
sub-categories, say Ia, Ib and IIa, IIb. Homleid (1995)
has some applications of this sort to meteorology,
where weather conditions were classified into such
categories and subcategories; her Markov models
had memory length three days. Model (6.1)
says that transitions from I to II take place 
at a rate $\alpha$ independent of the sub-categories,
and similarly from II to I at a rate $\beta$, 
and furthermore that there is a correspondence 
between movements inside I and inside~II.~\square


\section
\centerline{\bf 7. Estimation outside model conditions} 

\hop
In this section we leave the firm assumption that the
parametric model employed is fully correct, and learn
what then happens to the ML, the PL and the QL methods. 
Our operating assumption is simply that there are 
certain underlying transition probabilities
$$\pi_{a,b}=\Pr\{X_i=b\midd X_{i-1}=a\}
  \quad {\rm for\ }a,b=1,\ldots,S, $$
not necessarily inside the parametric class of $p_{a,b}(\theta)$. 
We shall see that for each of ML, PL, QL, the estimation
strategies aim at certain well-defined `last false'
parameter values, those providing the best parametric
approximation to the real data generating mechanism,
according to suitable Kullback--Leibler-related distance
measures. An initial analysis of this kind was in 
Hjort and Omre (1994), but the following discussion 
is both more complete and includes the PL. 

\subsection
{\sl 7.1. Least false for the ML method.} From 
the likelihood (1.2) we find
$$n^{-1}\log\lik_n(\theta)=\sum_{a,b}{N_{a,b}\over n}\log p_{a,b}(\theta)
   \arr_p\sum_{a,b}\pi_a\pi_{a,b}\log p_{a,b}(\theta)=H_\ML(\theta), $$
for each $\theta$, where the $\pi_a$s are the stationary 
probabilities coming from the set of transition probabilities 
$\pi_{a,b}$. Under natural regularity conditions the maximiser
of the left hand side, i.e.~the ML estimator, must converge
to the maximiser $\theta_0$ of the limit function $H_\ML(\theta)$. 
But this is the same as the minimiser of the distance measure
$$d_\ML(\truth,\model)=\sum_a\pi_a
   \Bigl\{\sum_b\pi_{a,b}\log{\pi_{a,b}\over p_{a,b}(\theta)}\Bigr\}. $$
This is a weighted average of Kullback--Leibler distances,
say $d_a(\truth_a,\model_a)$ for the $a$th row of 
the $P$ matrix, namely from $\pi_{a,\cdot}$ to 
the $p_{a,\cdot}(\theta)$ model. 

\subsection
{\sl 7.2. Least false for the PL method.} From 
the PL likelihood (1.3), 
$$\eqalign{
n^{-1}\log\pl_n(\theta)
&=\sum_{a,b,c}{N_{a,b,c}\over n}
   \log\Bigl\{{p_{a,b}(\theta)p_{b,c}(\theta)\over p_{a,c}^{(2)}(\theta)}
   \Bigr\} \cr
&\arr_p 
H_\PL(\theta)=\sum_{a,b,c}\pi_a\pi_{a,b}\pi_{b,c}
   \log\Bigl\{{p_{a,b}(\theta)p_{b,c}(\theta)\over p_{a,c}^{(2)}(\theta)}
   \Bigr\}. \cr}$$
Again, the PL estimator, which maximises the left hand side, will
under regularity conditions converge to the $\theta_0$ that maximises
the limit function $H_\PL(\theta)$. But this may be seen to be 
the same as minimising 
$$d_\PL(\truth,\model)=\sum_{a,c}\pi_a\pi_{a,c}^{(2)}
   \Bigl\{\sum_b{\pi_{a,b}\pi_{b,c}\over \pi_{a,c}^{(2)}}
   \log{\pi_{a,b}\pi_{b,c}/\pi_{a,c}^{(2)}
        \over p_{a,b}p_{b,c}/p_{a,c}^{(2)}}\Bigr\}. $$
This is once more in the form of a weighted sum of 
Kullback--Leibler distances, say $d_{a,c}(\truth_{a,c},\model_{a,c})$,
namely from the real model for $X_i$ given $(X_{i-1},X_{i+1})=(a,c)$
to the parametric model for the same distribution. 
We also note that an alternative perspective on the PL
method, stemming from (4.1), gives 
$$H_\PL(\theta)=2H_\ML(\theta)
   -\sum_{a,c}\pi_a\pi_{a,c}^{(2)}\log p_{a,c}^{(2)}(\theta), $$
where $\pi_{a,b}^{(2)}$ are the two-step probabilities
stemming from the matrix of the true $\pi_{a,b}$s. 
This is actually not a good property for the PL,
since good model fitting would mean both 
large value of $H_\ML$ and a large value of 
$\sum_{a,c}\pi_a\pi_{a,c}^{(2)}\log p_{a,c}^{(2)}(\theta)$,
and these desiderata are not well combined here;
cf.~the remarks made at the end of Sections 3 and 4,
and see the illustration below. 

\subsection 
{\sl 7.3. Least false for the QL method.} From (3.1), 
$$n^{-1}\log\ql_{n,k}(\theta)\arr_p
  H_\QL(\theta)=\sum_a\pi_a\log p_a(\theta)
  +(k-1)\sum_{a,b}\pi_a\pi_{a,b}\log p_{a,b}(\theta). $$
One finds that the QL estimator aims at the maximiser of 
the limit function, which is equivalent to minimising 
$$d_\QL(\truth,\model)=\sum_a\pi_a\log{\pi_a\over p_a(\theta)}
   +(k-1)\sum_a\pi_a
   \Bigl\{\sum_b\pi_{a,b}\log{\pi_{a,b}\over p_{a,b}}\Bigr\}. $$
This is in the form of a sum of two distances; first 
the Kullback--Leibler distance from the $\pi_a$s to the $p_a(\theta)$s,
and then the factor $(k-1)$ times the $\pi_a$-weighted 
combination of the Kullback--Leibler distances $d_a$
that we met for the ML method. 

\def\ha{\hskip7pt}

\subsection
{\sl 7.4. Illustration: using a four-parameter model
when a six-parameter model is true.}
Assume that a Markov chain on the four states 
A, G, C, T in reality is governed by a six-parameter
Kimura type model of the form  
$$\pmatrix{
1-2\alpha-\gamma_1 &\gamma_1      &\alpha      &\alpha \cr
\delta_1      &1-2\alpha-\delta_1 &\alpha      &\alpha \cr 
\beta      &\beta      &1-2\beta-\gamma_2 &\gamma_2 \cr
\beta      &\beta      &\delta_2      &1-2\beta-\delta_2 \cr},
   \eqno(7.1)$$
but that the somewhat cruder four-parameter model (6.1),
which assumes $\gamma_1=\gamma_2$ and $\delta_1=\delta_2$,
is being used for estimation and inference. 
We shall more precisely study the effects on (6.1)-based 
estimation and inference methods, when model (7.1) holds,  
under the particular circumstances
$$\alpha=.03, \ha 
  \beta=.04, \ha
  \gamma_1=.13+\eps, \ha 
  \gamma_2=.13-\eps, \ha
  \delta_1=.14+\eps, \ha 
  \delta_2=.14-\eps, $$ 
when $\eps$ ranges from $-0.10$ to $0.10$. 
In the middle, i.e.~when $\eps=0$, this is a real 
(6.1) model, with parameters $(.03,.04,.13,.14)$
meant to be plausible for the type of DNA analysis 
experiments considered in Section 6. 

\centerline{\includegraphics[scale=0.39]{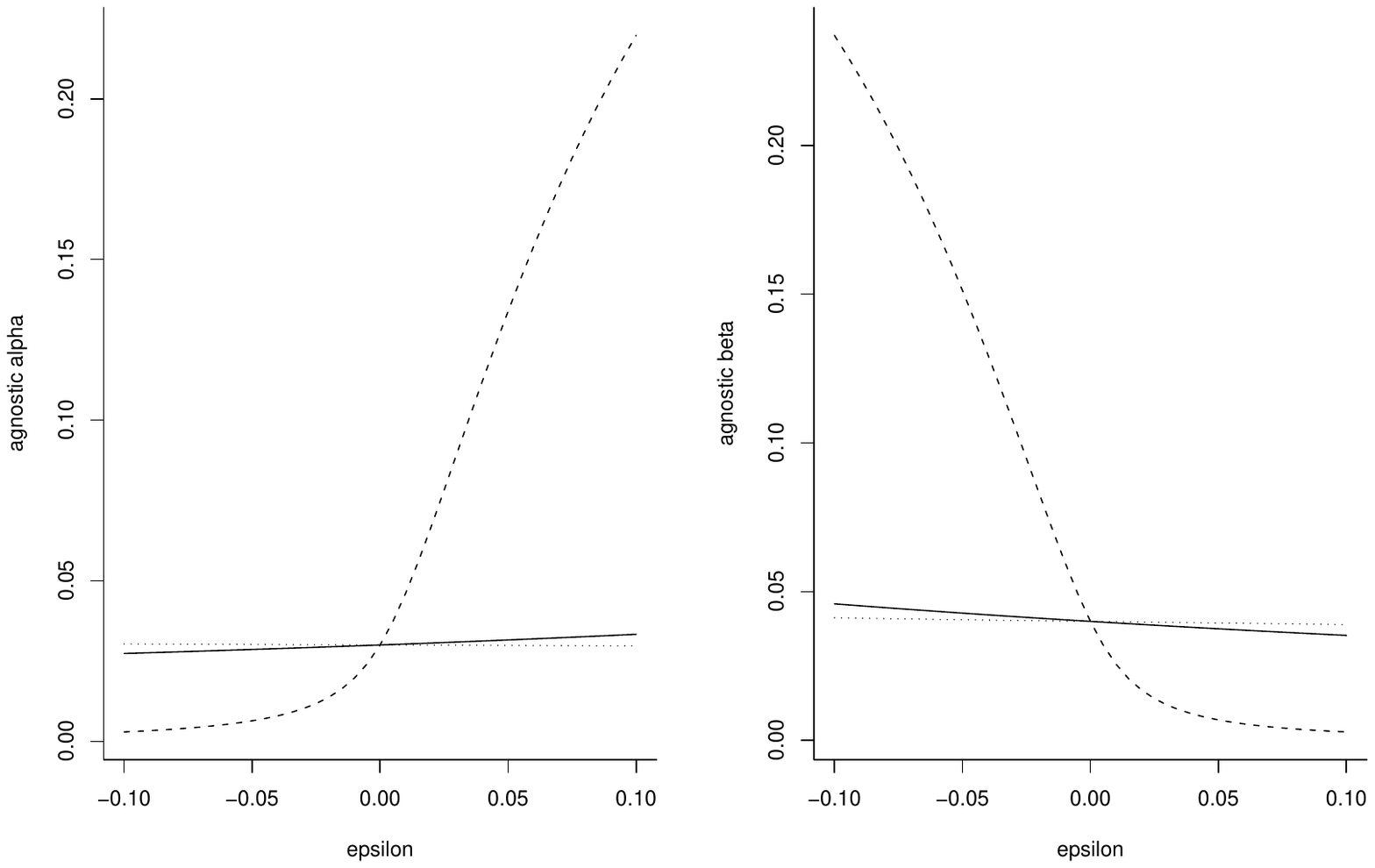}}

{\smallskip\narrower\sl\noindent\baselineskip11pt
{\csc Figure 7.1.}
The plot shows the least false parameter values,
for $\alpha$ and $\beta$, when the four-parameter 
Kimura model (6.1) is assumed, when the real mechanism 
is a six-parameter Kimura model, as a function 
of the model departure degree $\eps$; here 
$\gamma_1=\gamma+\eps$,
$\gamma_2=\gamma-\eps$,
$\delta_1=\delta+\eps$,
$\delta_2=\delta-\eps$,
and values $(.03,.04,.12,.14)$ are used for 
$(\alpha,\beta,\gamma,\delta)$. The least false
values are shown for the ML (solid curve),
the QL (dotted line, close to the ML),
and the PL (broken line). 
\smallskip}

\centerline{\includegraphics[scale=0.39]{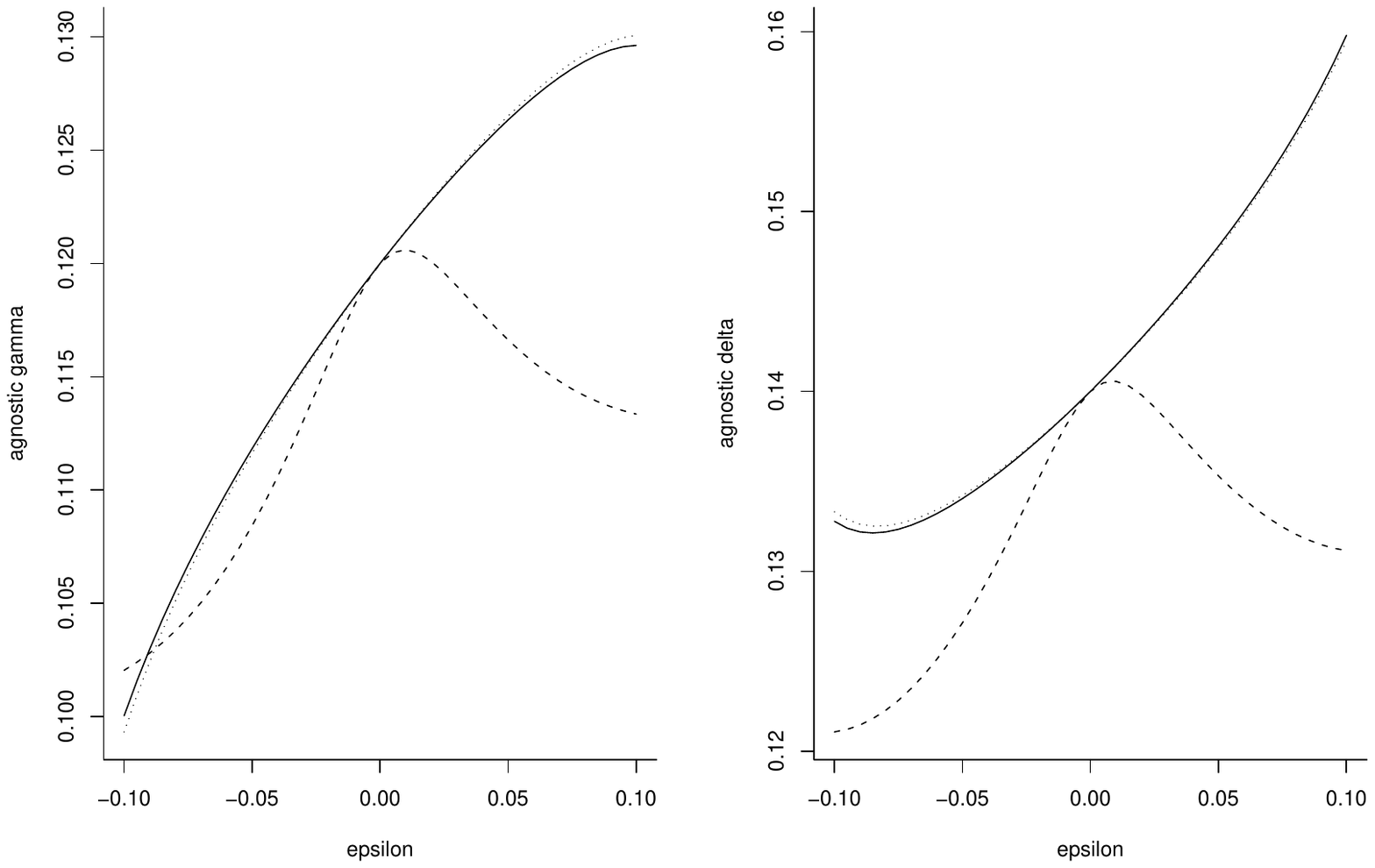}}

{\smallskip\narrower\sl\noindent\baselineskip11pt
{\csc Figure 7.2.}
The plot shows the least false parameter values,
for $\gamma$ and $\delta$, when the four-parameter 
Kimura model (6.1) is assumed, when the real mechanism 
is a six-parameter Kimura model, as a function 
of the model departure degree $\eps$; here 
$\gamma_1=\gamma+\eps$,
$\gamma_2=\gamma-\eps$,
$\delta_1=\delta+\eps$,
$\delta_2=\delta-\eps$,
and values $(.03,.04,.12,.14)$ are used for 
$(\alpha,\beta,\gamma,\delta)$. The least false
values are shown for the ML (solid curve),
the QL (dotted line, close to the ML),
and the PL (broken line). 
\smallskip}

For each $\eps$, (7.1) signifies a departure from the 
ideal model conditions, and the ML, QL, PL methods cope
with this in somewhat different ways, corresponding to
the different Kullback--Leibler weighting schemes
exhibited above.
Figures 7.1--7.2 show that the ML and
QL methods are in close agreement, and will aim at
estimating almost the same parameter values. 
The PL method reacts however rather differently 
to the model departure in question, and will aim for 
a different best-fitting four-parameter model. 
The difference in attitude stems from the PL perspective
of making it a priority to approximate all 
conditional probabilities 
$\Pr\{X_i=b\midd X_{i-1}=a,X_{i+1}=c\}$ well. 

\subsection
{\sl 7.5. Limiting normality.}
One may supplement the least false convergence results 
above with limiting normality statements, thus generalising
the ML, PL, QL propositions of earlier sections
from the `under the model' results to `outside the model' results. 
Such extensions are in principle not significantly 
more difficult to reach than the under the model results,
using the same techniques, but they become rather more complicated
in terms of notation. Thus we abstain from such efforts here.

\subsection
{\sl 7.6. Testing the models.}
Methods of the present section may be used
to check model adequacy by comparing variances
computed at the model and outside the model. 


\section
\centerline{\bf 8. Concluding remarks} 

\hop

\subsection
{\sl 8.1. Two large-sample frameworks.} 
Our framework for reaching large-sample results 
has been that of one long uninterrupted chain,
as in e.g.~Billingsley (1961a). In various
applications one would rather meet a growing
number of shorter chains, the situation essentially 
dealt with in Anderson and Goodman (1957). 
This does affect asymptotics in cases where 
the growing number of chains can be assumed
to start in their equilibrium distributions,
as the likelihoods would have a significant
factor, say $\prod_a p_a(\theta)^{O_a}$, 
with $O_a$ chains observed to have started in $a$. 
Without such an assumption the start information
is however either absent or negligible, and
the results and formulae we have derived remain valid.  
Among our reasons to work with the long chains 
rather than with many short ones here is that 
the former framework is closer to that met in
many spatial statistics applications. 

\subsection
{\sl 8.2. Markov chains with longer memory.} 
A Markov chain on $\{1,\ldots,S\}$ with 
memory length two steps rather than one step
can be transformed to a one-step memory 
Markov chain in the bigger state space 
$\{1,\ldots,S\}^2$. Hence methods and results
of our paper apply also to such models. 

\subsection
{\sl 8.3. Markov chain regression.} 
Our methods and results can under suitable conditions 
be generalised to situations where the transition
probabilities might depend on covariates. As 
a simple illustration, consider the three-state 
model of Section 5.4, and assume 
$$\alpha_i={\exp(r+sz_i)\over 1+\exp(r+sz_i)+\exp(t+uz_i)}, \quad
  \beta_i={\exp(t+uz_i)\over 1+\exp(r+sz_i)+\exp(t+uz_i)}, $$
in terms of a covariate $z_i$ thought to influence 
the $p_{a,b}$s at time $i$. One may then 
put up likelihood, pseudo-likelihood and quasi-likelihood
functions in terms of the parameters $(r,s,t,u)$,
and estimate via maximisation. 
Applications arise under either of the two scenarios
described in the preceding remark. 
Fokianos and Kedem (2003) operate with models of this sort,
using the ML method, and apply them to sleep data
and to DNA analysis. We expect that the QL method
will continue to perform well. 


\subsection
{\sl 8.4. Using QL for modelling purposes.} 
Markov chains have been propelled for about a 
hundred years by the modelling and analysis 
of transition probabilities, i.e.~of properties 
of $X_i$ given $X_{i-1}$. The good performance 
of the QL method may however invite further 
model constructions, where the basic motivation 
would come from joint modelling of $(X_{i-1},X_i)$,
or of triplets or quadruplets, instead of 
the conditional probabilities. A simple illustration
could take the form 
$$f_{a,b,c}=A(\rho)\exp[-\rho\{h(b,a)+h(b,c)\}]
  \quad {\rm for\ }a,b,c=1,\ldots,S $$ 
for the distribution of $(X_{i-1},X_i,X_{i+1})$, 
where $h(b,a)$ is a function being small for
$b$ `closer' to $a$ than when `farther away' from $a$;
examples include $h(b,a)=I\{b=a\}$ 
and $h(b,a)=|b-a|$. The $h$ function could also
be more elaborate, engaging further parameters. 
Transforming such models to transition probabilities,
for the Markov chain of appropriate oder, 
and then carry out inference, would be quite
cumbersome. But the QL method is well suited
to the task, starting from 
$$\sum_{i=2}^{n-1}\bigl[\log A(\rho)
   -\rho\{h(x_{i-1},x_i)+h(x_{i+1},x_i)\}\bigr]. $$

\subsection
{\sl 8.5. The QL in 2D.}  
For two-dimensional Markov fields, Besag (1974, 1975, 1977) 
invented the PL method, taking the product of 
all conditional probabilities given neighbourhoods, 
say $\prodin p_\theta(x_i\midd x_{\dell i})$,
where $\dell i$ denotes the set of neighbouring
sites to $i$. The QL method is cumbersome for Markov fields,
since computing the marginals of say $3\times3$ windows
is hard. But as for the preceding point 
the good properties of the QL might 
encourage invention of new spatial models that 
are better suited to computation of say 
$$\ql(\theta)=\prodin p_\theta(x_i,x_{\dell i})
   =\prodin p_\theta(x_i)p_\theta(x_{\dell i}\midd x_i). \eqno(8.1)$$
This is rather like turning the PL idea inside out. 
One particular class of models, for which 
the full likelihood is forbidding but (8.1) might
be worked with, is that of normal truncation models,
where one envisages that observations $x_i=h(z_i)$
for some thresholding transformation and where 
the $\{z_i\}$ is a Gau\ss ian process, for example
taken to be stationary and isotropic. A simple
random lattice example is $h(z)=I\{z>z_0\}$, 
and a covariance function say $K_\theta(\|z-z'\|)$ 
for the $z$ field. Here both ML and PL encounter 
serious problems, while (8.1) may be used 
to estimate $\theta$ and $z_0$. This idea
has actually been worked with 
by Heagerty and Lele (1998). They applied only
the pairwise version of the QL, however. QL versions
with triples or with smaller neighbourhoods
are possible to implement, and would typically
have stronger performance. 

As an indication of what can be worked with
along such lines, we now exhibit a binary lattice
process that can be thought of as similar to
the Gibbs process, but with different properties,
and with the QL being the natural estimation strategy.
Assume $\{z_i\}$ is a hidden stationary zero-mean 
Gau{\ss}ian process on a lattice, with covariance
function $K_\theta(\|i-j\|)$, perhaps having a white
noise component as the processes of Heagerty and Lele (1998)
have. Then model the observed binary process
$X_i$ as $I\{z_i>0\}$. Inference for the spatial
parameter $\theta$ could then take the QL 
of order corresponding to the four nearest neighbours;
$$\ql_n(\theta)=\prodin p_\theta(x_i,x_{i,N},x_{i,E},x_{i,S},x_{i,W}). $$
This is feasible with separately programmed 
functions for the 32 probabilities that 
a five-dimensional normal falls in the respective quintotants.


\subsection
{\sl 8.6. QL saves Neyman--Scott.}
An an application illustrating that QL can actually
be wiser than ML, consider the celebrated Neyman--Scott
problem, where an increasing number of nuisance 
parameters effectively destroys ML precision.
Here one observes independent pairs
$$\pmatrix{X_i \cr Y_i \cr}\sim\N_2(\pmatrix{\mu_i \cr \mu_i \cr},
  \pmatrix{\sigma^2 &0 \cr 0 &\sigma^2 \cr}) 
  \quad {\rm for\ }i=1,\ldots,n, $$
and one finds 
$$\hatt\sigma^2_\ML={1\over 4n}\sumin(X_i-Y_i)^2, $$ 
which travels to $\half\sigma^2$ rather than $\sigma^2$. 
The data information can be equivalently represented as 
$$\pmatrix{(X_i+Y_i)/2 \cr X_i-Y_i \cr}\sim\N_2
   (\pmatrix{\mu_i \cr 0 \cr},
  \pmatrix{\sigma^2/2 &0 \cr 0 &2\sigma^2 \cr}) 
  \quad {\rm for\ }i=1,\ldots,n, $$
and the natural QL strategy is to form a likelihood
using only the second halves, i.e.~the differences.
This gives a $\ql_n(\sigma)$ likelihood with maximiser  
$\hatt\sigma^2_\QL=2\hatt\sigma^2_\ML$, which is the right choice. 

\subsection
{\sl 8.7. General composite likelihoods.}
Finally, we point out that the term composite
likelihood appears to stem from Lindsay (1988),
who used it to indicate any function of parameters 
and data built by composition of valid 
likelihood objects. In this sense, ML, PL and QL 
are all examples of composite likelihoods.
The QL, for example, is the exact full likelihood
for an `imagined dataset', consisting in distinctly
placed $(X_{i-1},X_i)$ pairs. 

\section
\centerline{\bf Appendix: covariances for triplet counters} 

\hop
Here we find the covariance structure among the 
$S^3$-vector of $Z_{a,b,c}$s of the variables 
$Z_{n,a,b,c}=\rootn(N_{a,b,c}/n-p_ap_{a,b}p_{b,c})$. This was needed 
for Section 4. It follows as in Basawa and Rao (1980, Ch.~4)
that the $Z_{a,b,c}$s must be jointly normal with zero mean,
and that ${\rm cov}(Z_{a,b,c},Z_{d,e,f})$
must be equal to the limit of the covariance between
$Z_{n,a,b,c}$ and $Z_{n,d,e,f}$, as $n$ grows. 
This is the same as 
$$n^{-1}\sum_{i,j}{\rm cov}(I\{X_i=a,X_{i+1}=b,X_{i+2}=c\},
   I\{X_j=d,X_{j+1}=e,X_{j+2}=f\}), $$
where the sum extends over $i,j$ among $1,\ldots,n-2$. 
In fact one finds
$$\eqalign{{\rm cov}(Z_{a,b,c},Z_{d,e,f})
&=p_ap_{a,b}p_{b,c}(\delta_{a,d}\delta_{b,e}\delta_{c,f}-p_dp_{d,e}p_{e,f}) \cr
&\quad +p_ap_{a,b}p_{b,c}(\delta_{b,d}\delta_{c,e}-p_dp_{d,e})p_{e,f} \cr 
&\quad +p_dp_{d,e}p_{e,f}(\delta_{e,a}\delta_{f,b}-p_ap_{a,b})p_{b,c} \cr 
&\quad +p_ap_{a,b}p_{b,c}\gamma_{c,d}p_{d,e}p_{e,f}
  +p_dp_{d,e}p_{e,f}\gamma_{f,a}p_{a,b}p_{b,c}. \cr} \eqno({\rm A.1})$$
We reach result (A.1) by identifying and working with
different contributions from the double sum above. 

One contribution comes from $j=i$ terms, which gives
an expression with a limit recognisable as the first term of (A.1). 
Next, considering the contribution of $j=i+1$ terms
one finds the second term of (A.1), while 
similarly the third term of (A.1) comes from working
with $j=i-1$ terms and going to the limit. 
Then go on to terms where $j\ge i+2$, for which the
covariance in question may be expressed as 
$p_ap_{a,b}p_{b,c}(p_{c,d}^{(j-i-2)}-p_d)p_{d,e}p_{e,f}$.
Taking the limit as $n$ grows, and using (2.4), 
we find the fourth term of (A.1). Finally the fifth
term of (A.1) comes from the $i\ge j+2$ terms in the double sum. 

We take time to point out that result (2.3), 
which can be found in Basawa and Rao (Ch.~4), 
agrees with (A.1) in that we may sum (A.1) over all $c,f$
to find the covariance of $Z_{a,b}$ with $Z_{d,e}$. 
For Section 4 we also needed the following, 
obtained by summing over $b,e$: 
$$\eqalign{{\rm cov}(Z_{a,\cdot,c},Z_{d,\cdot,f})
&=p_ap_{a,c}^{(2)}(\delta_{a,d}\delta_{c,f}-p_dp_{d,f}^{(2)}) \cr
&\quad +p_ap_{a,d}p_{d,c}p_{c,f}+p_dp_{d,a}p_{a,f}p_{f,c}
        -2p_ap_{a,c}^{(2)}p_dp_{d,f}^{(2)} \cr 
&\quad +p_ap_{a,c}^{(2)}\gamma_{c,d}p_{d,f}^{(2)}
+p_dp_{d,f}^{(2)}\gamma_{f,a}p_{a,c }^{(2)}. \cr} \eqno({\rm A.2})$$

\bigskip
\centerline{\bf References}

\def\ref#1{{\noindent\hangafter=1\hangindent=20pt
  #1\smallskip}}          
\parindent0pt
\baselineskip11pt
\parskip3pt 
\medskip 

\ref{%
Anderson, T.W.~and Goodman, L.A. (1957).
Statistical inference about Markov chains.
{\sl Annals of Mathematical Statistics} {\bf 28}, 89--110.}


\ref{%
Barry, D.~and Hartigan, J.A. (1987).
Asynchronous distance between homologous DNA sequences.
{\sl Biometrics} {\bf 43}, 261--276.}

\ref{%
Basawa and Rao (1980).
{\sl Statistical Inference for Stochastic Processes.}
Academic Press, London.}

\ref{%
Basharin, G.P., Langville, A.N.~and Naumov, V.A. (2004).
The life and work of A.A.\allowbreak~Markov.
{\sl Linear Algebra and its Applications} {\bf 386}, 3--26.}

\ref{%
Besag, J. (1974). 
Spatial interaction and the statistical analysis of lattice
systems (with discussion contributions).
{\sl Journal of the Royal Statistical Society} {\bf B 36}, 192--236.}

\ref{%
Besag, J. (1975).
Statistical analysis of non-lattice data.
{\sl The Statistician} {\bf 24}, 179--195.}

\ref{%
Besag, J. (1977).
Some methods of statistical analysis for spatial data.
{\sl Bulletin of the Institute of International Statistics} {\bf 47}, 
77--92.} 

\ref{%
Billingsley, P. (1961a).
{\sl Statistical Inference for Markov Processes.}
Statistical Research Monographs, Vol.~II,
University of Chicago Press.} 

\ref{%
Billingsley, P. (1961b).
Statistical methods in Markov chains.
{\sl Annals of Mathematical Statistics} {\bf 32}, 12--40.}

\ref{%
Blaisdell, B. E. (1985). 
A method for estimating from two aligned
present day DNA sequences their ancestral composition and
subsequent rates of composition and subsequent rates of
substitution, possibly different in the two lineages, corrected
for multiple and parallel substitutions at the same site.
{\sl Journal of Molecular Evolution} {\bf 22}, 69--81.}

\ref{%
Cox, D. R.~and Reid, N. (2004).
A note on pseudolikelihood constructed from
marginal densities.
{\sl Biometrika} {\bf 91}, 729--737.}

\ref{%
Davison, A.C. (2003).
{\sl Statistical Models.}
Cambridge University Press, Cambridge.} 

\ref{%
Durret, R. (2002).
{\sl Probability Models for DNA Sequence Evolution.}
Probability and Its Applications,
Springer.} 

\ref{%
Fearnhead, P.~and Donnelly, P. (2002).
Approximate likelihood methods for estimating local recombination rates.
{\sl Journal of the Royal Statistical Society} 
{\bf B 64}, 657--680.}


\ref{%
Fokianos, K.~and Kedem, B. (2003).
Regression theory for categorical time series.
{\sl Statistical Science} {\bf 18}, 357--375.}


\ref{%
Glasbey, C.A. (2001).
Non-linear autoregressive time series with multivariate
Gaussian mixtures as marginal distributions.
{\sl Applied Statistics} {\bf 50}, 143--154.} 

\ref{%
Heagerty, P.J.~and Lele, S.R. (1998). 
A composite likelihood approach to binary spatial data.
{\sl Journal of the American Statistical Association} 
{\bf 93}, 1099--1111.}

\ref{%
Henderson, R. and Shimakura, S. (2003).
A serially correlated gamma frailty model for longitudinal count data.
{\sl Biometrika} {\bf 90}, 355--366.}

\ref{%
Hjort, N.L.~and Mohn, E. (1987).
Topics in the statistical analysis of remotely
sensed data [with discussion]. 
{\sl Bulletins of the International Statistical Institute} 
{\bf 52} (Proceedings of the ISI Meeting, Tokyo), 23--44.} 

\ref{%
Hjort, N.L.~and Omre, H. (1994).
Topics in spatial statistics (with discussion contributions).
{\sl Scandinavian Journal of Statistics} {\bf 21}, 289--357.}

\ref{%
Hjort, N.L.~and Mostad, P. (1998).
A quasi-likelihood method for estimating parameters 
in spatial covariance functions. Manuscript.} 

\ref{%
Hobolth, A.~and Jensen, J.L. (2005).
Statistical inference in evolutionary models
of DNA sequences via the EM algorithm.
Research report No.~455, Department of Theoretical Statistics,
University of Aarhus.}

\ref{%
Homleid, M. (1995).
{\sl Statistical Methods Applied in Meteorology.}
Cand.~scient.~thesis, Department of Mathematics,
University of Oslo.} 

\ref{%
Karlin, S.~and Taylor, H.M. (1975).
{\sl A First Course in Stochastic Processes.}
Academic Press, New York.}

\ref{%
Kimura, M. (1980).
A simple method for estimating evolutionary rates 
of base substitutions through comparative studies
of nucleotide sequences.
{\sl Journal of Molecular Evolution} {\bf 16}, 111--120.}

\ref{%
Kimura, M. (1981).
Estimation of evolutionary distances between 
homologous nucleotide sequences.
{\sl Proceedings of the National Academy of Sciences USA} 
{\bf 78}, 454--458.} 


\ref{%
de Leon, A.R. (2004).
Pairwise likelihood approach to grouped continuous model
and its extension. 
Technical report, Department of Mathematics \& Statistics,
University of Calgary.} 

\ref{%
Lindsay, B. (1988).
Composite likelihood methods.
In {\sl Statistical Inference for Stochastic Processes}
(ed.~N.U.~Prahbu), American Mathematical Society.}


\font\cyr=wncyr10

\font\cyss=wncyss10
\def\j3{{\rm\u{\cyr i}}}

\ref{%
{\cyr
Markov, A.A. (1906).
Rasprostranenie zakona bolp1\-shih chisel na velichiny,
zavisyashchie drug ot druga.}
{\cyss Izvestiya Fiziko-matematicheskogo obchestva
pri Ka\-zanskom universitete} {\bf 15} {\cyr (2-ya seriya)},
124--156.}

\ref{%
{\cyr
Markov, A.A. (1913).
Primer statisticheskogo issledovaniya nad tekstom
``Evgeniya Onegina'', illyustriruyushchi{\j3} 
svyazp1 ispytani{\j3} v cepp1.
{\cyss Izvestiya Aka\-demii Nauk}, 
Sankt-Peterburg {\bf 7} (6-ya seriya), 153--162.}}

\ref{%
Nott, D.J.~and Ryd\'en, T. (1999).
Pairwise likelihood methods for inference in image models.
{\sl Biometrika} {\bf 86}, 661--676.}

\ref{%
Parner, E.T. (2001).
A composite likelihood approach to multivariate survival data.
{\sl Scandinavian Journal of Statistics} {\bf 28}, 295--302.}

\ref{%
Pickard, D.K. (1987).
Inference for discrete Markov fields:
the simplest nontrivial case.
{\sl Journal of the American Statistical Association}
{\bf 82}, 90--96.} 

\ref{%
Pushkin, A.S. (1977). 
{\sl Eugene Onegin} [translated by C.H.~Johnston].
Penguin Classics, London. 
There are various later reprints of essentially 
the same translation of Pushkin's 1833 epic.}

\ref{%
Renard, D., Molenberghs, G.~and Geys, H. (2004).
A pairwise likelihood approach to estimation in multilevel probit models.
{\sl Computational Statistics \& Data Analysis} {\bf 44}, 649--667.}

\ref{%
Strauss, D.~and Ikeda, M. (1990).
Pseudolikelihood estimation for social networks.
{\sl Journal of the American Statistical Association} 
{\bf 85}, 204--212.}

\ref{%
Taylor, H.M.~and Karlin, S. (1984).
{\sl An Introduction to Stochastic Modeling.}
Academic Press, New York.} 

\ref{%
Varin, C., H{\o}st, G.~and Skare, \O. (2005).
Pairwise likelihood inference in spatial generalized linear mixed models.
{\sl Computational Statistics \& Data Analysis}, to appear.}


\bye